\documentclass[11pt]{iopart}

\usepackage{graphicx}
\usepackage{amssymb}
\usepackage{exscale}
\usepackage{iopams}
\usepackage{harvard}
\usepackage{amsfonts}
\usepackage{epsfig}
\usepackage{psboxit}
\usepackage{xspace}
\usepackage{color}
\usepackage{placeins}
\usepackage{subfigure}
\usepackage{xspace}

\begin{document}
\renewcommand{\Im}[0]{\mathrm{Im}\,}
\renewcommand{\Re}[0]{\mathrm{Re}\,}

\renewcommand{\epsilon}{\varepsilon}

\newcommand{\ie}[0]{i.e.\@\xspace}
\newcommand{\eg}[0]{e.g.\@\xspace}

\newcommand{\om}[0]{\omega}
\newcommand{\gammab}[0]{\bar{\gamma}}
\newcommand{\Ep}{E_\mathrm{p}}
\newcommand{\omb}[0]{\bar{\omega}}
\newcommand{\Ap}[0]{A_\mathrm{p}}
\newcommand{\Ae}[0]{A_\mathrm{e}}
\newcommand{\me}[0]{\mathrm{e}}
\newcommand{\mi}[0]{\mathrm{i}}
\newcommand{\md}[0]{\mathrm{d}}
\newcommand{\nag}{\phantom{\dag}}
\newcommand{\op}{\hat{p}}
\newcommand{\ox}{\hat{x}}
\newcommand{\on}{\hat{n}}

\renewcommand{\comment}[1]{ {\bf !!! #1 !!! } }

\newcommand{\las}[0]{\langle}
\newcommand{\ras}[0]{\rangle}
\newcommand{\la}[0]{\left\las}
\newcommand{\ra}[0]{\right\ras}
\newcommand{\ket}[1]{\left|#1\ra}
\newcommand{\bra}[1]{\la#1\right|}
\newcommand{\sket}[1]{|#1\ras}  
\newcommand{\sbra}[1]{\las#1|} 
\newcommand{\braket}[2]{\la#1\left.\right|#2\ra}
\newcommand{\sbraket}[2]{\las#1\left.\right|#2\ras}

\newcommand{\dbraket}[2]{\la\la#1\left.\right|#2\ra\ra}

\renewcommand\floatpagefraction{.9}
\renewcommand\topfraction{.9}
\renewcommand\bottomfraction{.9}
\renewcommand\textfraction{.1}

\graphicspath{ {.} {./Bilder-090407-agrandeps/} }

\title{Phonon affected transport through molecular quantum dots}

\author{J Loos\dag, T Koch\ddag, A Alvermann\ddag, A R Bishop\S,  
and H Fehske\ddag}

\address{\dag\ %
  Institute of Physics, Academy of Sciences of the Czech Republic, 16200
Prague, Czech Republic
}
\address{\ddag\ %
  Institute of Physics, Ernst-Moritz-Arndt University Greifswald, 
17487 Greifswald, Germany
}
\address{\S\ %
  Theory, Simulation and Computation Directorate, 
Los Alamos National Laboratory, Los Alamos, New Mexico 87545, USA
}
\ead{\mailto{loos@fzu.cz}}

\begin{abstract}
To describe the interaction of molecular vibrations with electrons at 
a quantum dot contacted to metallic leads, we extend an analytical 
approach that we previously developed for the many-polaron problem.
Our scheme is based on an incomplete variational Lang-Firsov transformation,
combined with a perturbative calculation of the electron-phonon self-energy 
in the framework of generalised Matsubara functions.
This allows us to describe the system at weak to strong coupling and 
intermediate to large phonon frequencies. 
We present results for the quantum dot spectral function and 
for the kinetic coefficient that characterises the electron transport through 
the dot. With these results we critically examine the strengths 
and limitations of our approach, and discuss the properties of the molecular quantum dot in the context of polaron physics.
We place particular emphasis on the importance of corrections 
to the concept of an antiadiabatic dot polaron suggested by the 
complete Lang-Firsov transformation.

\end{abstract}

\pacs{73.63.Kv, 71.38.-k, 73.21.La, 72.10.-d}

\submitto{\JPCM}


\section{Introduction}\label{sec:introduction}
Recent advances in nanotechnology have stimulated great interest
in the basic mechanisms of transport through molecular 
junctions~\cite{CRRT90,Paea02,ROBWML02,KDHCBSHB03,Pa07,CFR05}. In such devices the central element 
can be a single organic molecule or a suspended carbon nanotube, 
which may be thought of as a quantum dot contacted  
to metallic leads that act as macroscopic charge reservoirs. 
Transport through such a quantum dot is determined by energy level quantisation as well as electronic correlations and electron-phonon (EP) interaction~\cite{AB03,MAM04,TKM05,GRN07,FWLB08}.

Vibrations of a molecular quantum dot are 
local excitations of substantial energy, which are represented by 
optical phonons. Their frequency 
is comparable to the transfer integral or kinetic energy of electrons~\cite{NCUB07}. Therefore, the mobility of electrons 
is significantly modified by the influence of molecular vibrations. 
In this respect, a molecular quantum dot resembles the situation 
in a crystalline structure, where the coupling between vibrations and electrons may lead to the formation of (small) polarons, as studied 
in the context of Holstein's molecular crystal model~\cite{Ho59a,Ho59b}.
A Holstein polaron is an electron dressed by a phonon cloud. 
Since the polaron must carry the accompanying deformation through the lattice, 
the mobility of Holstein polarons can be renormalised by several orders of magnitude in comparison to the free electronic excitation~\cite{LF62,WF98}.  While the physics of Holstein polarons in a perfect crystal at low temperature 
and small density is by now well understood (see e.g. the 
review ~\cite{FT07}), there is less understanding if the periodicity 
of the crystal is altered, e.g. by impurities~\cite{MNAFDCS09,AF08b} 
or disorder~\cite{BF02,BAF04}, in anisotropic materials~\cite{AFT08,Emi86}, 
or is absent for complicated geometries.

For a molecular quantum dot, translational symmetry is broken from the outset,  and EP coupling is relevant only in a small part of the entire system. 
The electron current through a deformable quantum dot was found 
to depend significantly on the local EP 
coupling~\cite{Fl03,NCUB07,ZM07,MAM04,TKM05,HF07}. 
In order to understand the basic transport mechanisms in such devices,  
appropriate theoretical models have to be studied.
The most simple model corresponds to a modified Fano-Anderson model, 
where a vibrating quantum dot replaces the static 
impurity.  Then the current is determined 
by the dot spectral function~\cite{MW92a}.
The spectral function accounts for the leads, 
as well as for the influence of EP coupling.
In particular, it determines the charge carrier population of the dot,
and its value close to the Fermi energy of the leads 
determines the number of electrons contributing to the current.
Since, with increasing EP interaction, spectral weight is transferred to 
lower energies, the charge carrier population of the dot increases.
For the current, on the other hand, a reduction is expected since 
the spectral weight at the Fermi energy decreases.

In the present manuscript we will address electron transport through 
a deformable molecule within an approximate description, which we 
previously developed for Holstein polarons at finite 
density~\cite{LHF06,LHAF06,LHAF07}.
It accounts for renormalisation of transport and inelastic processes, and Pauli blocking. Higher order many-particle processes, namely the further excitation of electron-hole pairs and subsequent evolution of many-particle correlations, 
are not included.
The current presentation, therefore, should be considered as an 
important but intermediate step towards a complete description.
A particular feature of our approach is that it interpolates between weak and strong coupling using an incomplete variational Lang-Firsov transformation.
As a consequence it describes polaronic effects  without being restricted to
 the antiadiabatic strong coupling regime.
We introduce our approach here for the current in linear response, where the kinetic coefficient is obtained from the dot spectral function at equilibrium.
Subsequent work will address the current at finite voltage bias.

The paper is organised as follows.
In section~\ref{sec:model} we introduce the model Hamiltonian, and describe the variational Lang-Firsov transformation.
In section~\ref{sec:specfunc} we derive the expressions and the iterative calculation scheme for the self-energy, which depends on the variational parameter of the incomplete Lang-Firsov transformation. This parameter is obtained from minimisation of the energy, which we express as the expectation value of the Hamiltonian
within the approximation used. 
From the spectral function, the kinetic 
coefficient is obtained in section~\ref{sec:kincoeff}.
Section~\ref{sec:results} discusses the numerical results,
and we conclude in section~\ref{sec:conclusion}.

\section{Theoretical Approach}\label{sec:theory}
\subsection{Model}\label{sec:model}
\begin{figure}
  \begin{center}
\includegraphics[width=0.6\textwidth]{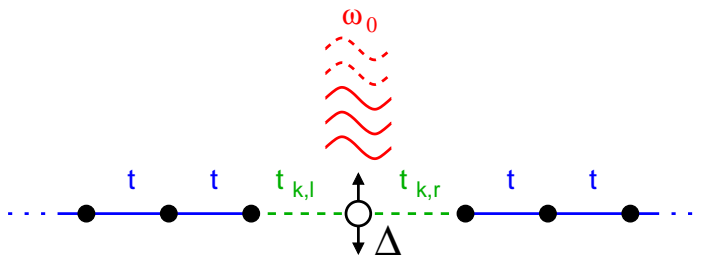}
  \end{center}
\caption{\label{hu_model}%
    (Colour online) Sketch of a molecular quantum dot with vibrational frequency $\omega_0$ between two metallic leads.}
\end{figure}

The paradigmatic example of a vibrating quantum dot is provided by 
a molecule sandwiched between two metallic leads 
(see figure~\ref{hu_model}). 
Such a system
can be described by the Hamiltonian
\begin{eqnarray}
\label{ho1}
   H  &=&  \sum_{k,a} (E_{ka}^{\nag}-\mu_a^{\nag})c_{ka}^{\dag} c_{ka}^{\nag} -\frac{t_d}{\sqrt{N}} \sum_{k,a} \left(d^{\dag}c_{ka}^{\nag}+c_{ka}^{\dag}d\right)\nonumber\\ 
&&+ (\Delta-\mu) d^\dag d^{\nag}  -g \om_0 ( b^{\dag} + b) d^\dag d  + \om_0  b^{\dag} b \,.
\end{eqnarray}
Here, the $E_{ka}$ (for $k=1,\dots,N$) give the energies of non-interacting electrons in the left and right lead $a=l,r$,
and $c_{ka}^{\dag}$ ($c_{ka}^{\nag}$) are the corresponding 
creation (destruction) operators of free fermions in the $N$ lead states. The leads will later be specified by their density of states $\varrho(\xi)=\frac{1}{N}\sum_k\delta[\xi-E_k]$; the population of the leads is determined by the chemical potentials $\mu_a$. 
The quantum dot is 
represented by a single energy level $\Delta$, with 
Fermi operators $d^{(\dag)}$ and the chemical potential $\mu=(\mu_l+\mu_r)/2$. The term $\propto t_{d}$ allows for lead-dot particle transfer; a possible $k$-dependence of the dot-lead-coupling can be absorbed in $\varrho(\xi)$.
An electron at the quantum dot interacts via a 
Holstein-type coupling with a local (intra-molecular) vibrational mode; 
$g$ denotes the dimensionless EP coupling constant,
and $\om_0$ the frequency of the optical
phonons created (annihilated) by $b^\dag$ ($b$).

The quantum dot responds to the presence of an electron with a finite deformation.
For sufficiently large phonon frequency $\omega_0$, the strength of the deformation depends only on the momentary occupancy of the dot.
This is in analogy to Holstein's small polaron theory,
where a lattice deformation in the vicinity of the electron 
accompanies the electron motion.
To describe this effect, we apply 
a generalised Lang-Firsov displacement transformation~\cite{LF62} with parameter $\gamma\in [0,1]$, 
\begin{equation}
\label{distrafo}
U  = \me^{\tilde g(b^{\dag}-b)d^{\dag}d}\;,
\quad {\rm for} \quad\quad \tilde g  =  \gamma g\;.
\end{equation}
After this transformation, the original electron and phonon operators are given as
\begin{equation}
\label{traop}
\tilde d =  \me^{\tilde g (b^{\dag}-b)}\,d \;,\quad \tilde b  =  b+\tilde g\, d^{\dag}d\;.
\end{equation}
The transformed Hamiltonian $\tilde H = U^\dagger H U$ reads  
\begin{eqnarray}
\label{ho2}
\tilde{H} &=&\sum_{k,a} (E_{ka}^{\nag}-\mu_a^{\nag}) c_{ka}^{\dag} c_{ka}^{\nag}
 - \sum_{k,a}\left(C^{\phantom{\nag}}_{t}d^{\dag}c_{ka}^{\nag}+C^{\dag}_{t}c_{ka}^{\dag}d\right)\nonumber\\
&&\quad+(\tilde \Delta-\mu) d^\dag d^{\nag} -C_{d}d^{\dag} d + \om_0     b^{\dag} b \;,
\end{eqnarray}
where
\begin{equation}\label{tr1}
C^{\nag}_{t}=\frac{1}{\sqrt{N}}t_{d}\,\me^{-\tilde g(b^{\dag}-b)} \,,\quad
	C^{\nag}_{d}=g\om_0(1-\gamma)(b^{\dag}+b)\,,	
\end{equation}
and
\begin{equation}
\label{tr2}
\tilde \Delta=\Delta-\varepsilon_{p}\gamma(2-\gamma)\;, \; {\rm with } \;\; \varepsilon_{p}=g^2\om_0\,.
\end{equation}
%
As the parameter $\gamma$ of the Lang-Firsov transformation 
grows from $\gamma=0$ to $\gamma=1$ it accounts for the transition between 
the weak-coupling and strong-coupling regimes.
The value of $\gamma$ will be later determined from minimisation of the energy.
Only for very strong coupling and large phonon frequency,
the value $\gamma=1$ is approached. Then, the canonical transformation~(\ref{distrafo}) eliminates the direct coupling term between the new fermion and shifted 
boson operators at the price of introducing a boson modified transfer term between quantum dot and leads. This corresponds to the strong-coupling limit of polaron theory, where the new Fermi operators $d$ would represent small polarons in the deformable lattice.
For our problem, `polaron formation' at the quantum dot mainly results in lowering of the dot energy level by the polaron shift $\varepsilon_p$, and in an exponential reduction of the effective dot-lead transfer $\tilde{t}_{d} = t_{d} e^{-\tilde{g}^2/2}$.
Note that the variation of $\gamma$ throughout the parameter regime
is important to describe the system away from the strong-coupling limit.
The use of the $\gamma$-dependent variational Lang-Firsov transformation is an essential feature of our description.

\subsection{Single-particle properties: quantum-dot 
spectral function}\label{sec:specfunc}

We first determine the retarded Green function $G^R_{dd}$ of the quantum dot, which is represented by the operators $d^{(\dag)}$ in the transformed Hamiltonian~(\ref{ho2}). 
The Green function is calculated within 
perturbation theory 
up to second order in the interaction coefficients~(\ref{tr1}), starting from the Lang-Firsov transformed Hamiltonian~(\ref{ho2}).
Since the parameter $\gamma$ is assigned variationally,
this treatment exceeds standard weak-coupling or strong-coupling   
perturbation theory,
which starts either from the untransformed Hamiltonian (corresponding to $\gamma=0$) or the fully transformed Hamiltonian ($\gamma=1$).
The combination of perturbation theory with an incomplete variational Lang-Firsov transformation provides meaningful results also away from these limiting cases.

Our calculation is based on the equations 
of motion for the generalised temperature Green functions~\cite{KB62},   
adapted to systems with EP interaction~\cite{BT62,Sc66}.
Accordingly we define 
\begin{equation}
\label{gdd}
G_{dd}(\tau_1,\tau_2;\{V\})=-\frac{1}{\langle S \rangle} \langle \mathcal{T}_{\tau}d(\tau_1)d^{\dag}(\tau_2)S  \rangle\,,
\end{equation}
and in an analogous way $G_{cd;ka}$, $G_{dc;ka}$, and $G_{cc;ka}$. The mean value and (imaginary) time dependencies in (\ref{gdd}) are determined by $\tilde H$ with $\mu=\mu_l=\mu_r$, the equilibrium chemical potential of the system. Moreover we set 
\begin{equation}\label{smatrix}
S  =  \mathcal{T}_{\tau}\exp\left\{-\int_{0}^{\beta}\mathrm{d}\tau\, V^{\nag}_{t}(\tau)C^{\nag}_{t}(\tau)+\bar{V}^{\nag}_{t}(\tau)C^{\dag}_{t}(\tau)+V^{\nag}_{d}(\tau)C^{\nag}_{d}(\tau)\right\}\,,
 \end{equation}
where $\beta$ is the inverse temperature, and the classical variables $V^{\nag}_{t}$, $\bar{V}^{\nag}_{t}$, and $V^{\nag}_{d}$ are introduced as a purely formal device. 

We set up the equations of motion for the Green functions using the following  
matrix notation
\begin{equation}
\label{matrixmult}
\int_{0}^{\beta}\mathrm{d}\tau' G_1(\tau_1,\tau';\{V\})G_2(\tau',\tau_2;\{V\}) \equiv  G_1(\tau_1,\tau';\{V\}) \circ G_2(\tau',\tau_2;\{V\})\,.
\end{equation}
If $G_1$, $G_2$ satisfy the relation  
\begin{equation}
G_1(\tau_1,\tau';\{V\}) \circ G_2 (\tau',\tau_2;\{V\})  =  \delta[\tau_1-\tau_2] \;,
\end{equation}
they are called inverse functions of each other. 
In particular, the inverse functions to the zeroth-order Green functions
$G_{dd}^{(0)}(\tau_1,\tau_2)$ and $G_{cc;\,ka}^{(0)}(\tau_1,\tau_2)$ are given as 
\begin{equation}
G_{dd}^{(0)-1}(\tau_1,\tau_2)  =  \left [ -\frac{\partial}{\partial\tau_1}-(\tilde\Delta-\mu) \right] \delta[\tau_1-\tau_2] 
\end{equation}
and
\begin{equation}
\label{Green_unpert_lead}
G_{cc;\,ka}^{(0)-1}(\tau_1,\tau_2) =  \left [ -\frac{\partial}{\partial\tau_1}-(E_{ka}-\mu) \right] \delta[\tau_1-\tau_2]\,,
\end{equation}
respectively. 
By functional derivation with respect to the 
auxiliary fields $\{V\}$ we find a set of coupled equations,
\begin{eqnarray}
&&G_{dd}^{(0)-1}(\tau_1,\tau')\circ G_{dd}(\tau',\tau_2;\{V\}) = \delta[\tau_1-\tau_2]\nonumber\\
&&\qquad-\bar{C}_d(\tau_1,\{V\})G_{dd}(\tau_1,\tau_2;\{V\})+\frac{\delta}{\delta V_d(\tau_1)}G_{dd}(\tau_1,\tau_2;\{V\})
\nonumber\\&&\qquad-\sum_{k,a}\bar{C}_{t}(\tau_1,\{V\})
G_{cd;\,ka}(\tau_1,\tau_2;\{V\})\nonumber\\
&&\qquad+\sum_{k,a}\frac{\delta}{\delta V_{t}(\tau_1)}G_{cd;\,ka}(\tau_1,\tau_2;\{V\})\label{g0dd-1}\,,
\end{eqnarray}
\begin{eqnarray}
&&G_{cc;\,ka}^{(0)-1}(\tau_1,\tau')\circ G_{cd;\,ka}(\tau',\tau_2;\{V\}) = \nonumber\\&&
\qquad-\bar{C}_{t}^{\dag}(\tau_1,\{V\})G_{dd}(\tau_1,\tau_2;\{V\})
+\frac{\delta}{\delta \bar{V}_{t}(\tau_1)}G_{dd}^{\nag}(\tau_1,\tau_2;\{V\})\,,
\label{g0cc-1}
  \end{eqnarray}
with  $\bar{C}_{d}(\tau,\{V\})=\frac{1}{\langle S\rangle} \langle \mathcal{T}_{\tau} C_d(\tau)S \rangle$, 
  $\bar{C}_{t}(\tau,\{V\})= \frac{1}{\langle S\rangle}\langle \mathcal{T}_{\tau} C_{t}(\tau)S \rangle$,
  and $\bar{C}_{t}^{\dag}(\tau,\{V\})= \frac{1}{\langle S\rangle}\langle \mathcal{T}_{\tau} C_{t}^\dag(\tau)S \rangle$.

In order to solve this system of equations, we multiply~(\ref{g0cc-1}) by 
$G_{cc;ka}^{(0)}$ from the left and substitute the resulting expression for
$G_{cd;ka}$ in~(\ref{g0dd-1}). Then equation~(\ref{g0dd-1}) is
multiplied by $G_{dd}^{-1}$ from the right. The resulting equation for
$G_{dd}$ is converted to an equation for the self-energy $\Sigma_{dd}$,
introduced by 
\begin{equation}
G_{dd}^{-1}(\tau_1,\tau_2;\{V\})  =  G_{dd}^{(0)-1}(\tau_1,\tau_2)-\Sigma_{dd} (\tau_1,\tau_2;\{V\})\,. 
\end{equation}
By use of the functional differentiation rules 
  \begin{equation}
\label{ functder}
\delta G\circ G^{-1}=-G\circ \delta G^{-1}=G\circ\delta\Sigma,\quad\delta G=G\circ\delta\Sigma\circ G\,,
\end{equation}
the self-energy $\Sigma$ becomes
  \begin{eqnarray}\label{sigmadiff}
& &\hspace*{-0.4cm} \Sigma_{dd} ( \tau_1,\tau_2;\{V\})  = - \bar{ C}_d (\tau_1;\{V\})  \delta[\tau_1-\tau_2] \\[0.2cm]&&  + \sum_{k,a} \bar{ C}_{t} (\tau_1;\{V\}) G_{cc;\,ka}^{(0)}(\tau_1,\tau_2) \bar{ C}_{t}^\dag (\tau_2;\{V\}) \nonumber \\
& & 	- \sum_{k,a}G_{cc;\,ka}^{(0)}(\tau_1,\tau_2) \frac{\delta\bar{ C}_{t}^\dag (\tau_2;\{V\}) }{\delta V_{t}(\tau_1) } 
	+ G_{dd}(\tau_1,\tau';\{V\}) \circ \frac{\delta\Sigma_{dd} ( \tau',\tau_2;\{V\})}{\delta V_{d}(\tau_1) }  \nonumber\\
& & 	- \sum_{k,a} \bar{ C}_{t} (\tau_1;\{V\}) G_{cc;\,ka}^{(0)}(\tau_1,\tau'') \circ G_{dd}(\tau'',\tau';\{V\}) \circ
		\frac{\delta\Sigma_{dd} ( \tau',\tau_2;\{V\}) }{\delta \bar{V}_{t} (\tau'') }  
  \nonumber\\
& & 	- \sum_{k,a} G_{cc;\,ka}^{(0)}(\tau_1,\tau'') \bar{ C}_{t}^\dag (\tau'';\{V\})  \circ G_{dd}(\tau'',\tau';\{V\}) \circ
		\frac{\delta\Sigma_{dd} ( \tau',\tau_2;\{V\})}{\delta V_{t}(\tau_1) }   \nonumber\\ 
& & 	+ \textnormal{ terms with products of functional derivatives of $\Sigma_{dd}$}
  \nonumber\\&&+\textnormal{
  terms with second functional derivatives of $\Sigma_{dd}$.}\nonumber 
\end{eqnarray}
Within our iterative scheme, the terms on the r.h.s. of equation~(\ref{sigmadiff}) 
without functional derivatives of $\Sigma_{dd}$ are taken in the first step as
$\Sigma_{dd}^{(1)} ( \tau_1,\tau_2;\{V\})$. Explicitly, we have 
  \begin{eqnarray}\label{sigmaone}
&&\hspace*{-0.4cm}\Sigma_{dd}^{(1)} ( \tau_1,\tau_2;\{V\}) =  
  - \bar{ C}_d (\tau_1;\{V\}) \delta[\tau_1-\tau_2]\\[0.3cm]
& & + \sum_{k,a} \bar{ C}_{t}^{\nag} (\tau_1;\{V\}) G_{cc;\,ka}^{(0)}(\tau_1,\tau_2) \bar{ C}_{t}^\dag (\tau_2;\{V\}) \nonumber\\
& & +\sum_{k,a}G_{cc;\,ka}^{(0)}(\tau_1,\tau_2) 
  \left[ \frac{1}{\langle S\rangle}\langle \mathcal{T}_\tau C_{t}^{\nag}(\tau_1) C_{t}^{\dag}(\tau_2) S\rangle - \bar{C}_{t}^{\nag}(\tau_1;\{V\})\bar{C}_{t}^{\dag} (\tau_2;\{V\})\right].\nonumber
\end{eqnarray}
In the second step, to obtain $\Sigma_{dd}^{(2)}$, the first approximation
$\Sigma_{dd}^{(1)}$ has to be inserted into the functional derivatives
of $\Sigma_{dd}$ on the r.h.s. of~(\ref{sigmadiff}). But, if we confine 
ourselves to terms up to second order in the interaction coefficients, only the 
first four terms on the r.h.s. are relevant. In this approximation,  
the self-energy 
$\Sigma_{dd}(\tau_1,\tau_2)=\Sigma_{dd}(\tau_1,\tau_2;\{0\})$, determining the
temperature Green function 
$G_{dd}(\tau_1,\tau_2)=G_{dd}(\tau_1,\tau_2;\{0\})$, is given as  
\begin{equation}
\label{sdd}
\Sigma_{dd}^{\nag}(\tau_1,\tau_2)=\Sigma_{dd}^{(1)}(\tau_1,\tau_2;\{0\})+
G_{dd}^{\nag}(\tau_1,\tau_2)\langle \mathcal{T}_\tau 
C_d(\tau_1)C_d(\tau_2)\rangle\,,
\end{equation}
where
$\bar{C}_d(\tau;\{0\})=0$ and 
$\bar{C}_{t}^{\nag}(\tau;\{0\})=[\bar{C}_{t}^{\dag}(\tau;\{0\})]^*=\frac{1}{\sqrt{N}}t_{d}\exp\{-\frac{1}{2}\tilde{g}^2\coth (\frac{1}{2}\beta\omega_0)\}$.
The correlation functions of the interaction coefficients
occurring in~(\ref{sdd}) have been calculated previously for the
generalised Lang-Firsov transformation in the Holstein model~\cite{FLW97}.
Converting~(\ref{sdd}) to the equation for the Fourier transform of
the self-energy and expressing the Fourier transform of $G_{dd}$,
$G_{dd}(\mi \omega_{\nu})$, by means of the spectral function 
$A_{dd}$, the summation over the bosonic Matsubara frequencies 
$\omega_{n}= 2 n \pi/\beta$ (being the difference of two fermionic 
Matsubara frequencies) can be carried
out, and we obtain, within low-temperature approximation 
$\beta\omega_0\gg 1$,   
\begin{eqnarray} \label{sigmatwocomplex}
&&\Sigma_{dd}(\mi\om_{\nu})=\frac{1}{N}\,t_d^2\,\me^{-\tilde g^2}\sum_{k,a}\frac{1}{\mi\om_{\nu}-(E_{ka}-\mu)} \\
	&&\quad + \sum_{s\geq 1}\me^{-\tilde g^2}\frac{(\tilde g^2)^{s}}{s!}\frac{1}{N}\,t_d^2\sum_{k,a}\left(\frac{n_{F}(E_{ka}-\mu)}{i\om_{\nu}-(E_{ka}-\mu)+s\om_0}
	+ \frac{1-n_{F}(E_{ka}-\mu)}{\mi\om_{\nu}-(E_{ka}-\mu)-s\om_0}\right) \nonumber\\
	&&\quad +\left [(1-\gamma)g\om_0\right ]^2\int_{-\infty}^{+\infty}\mathrm{d}\om'A_{dd}(\om')\left(\frac{n_{F}(\om')}{i\om_{\nu}-\om'+\om_0}+
	\frac{1-n_{F}(\om')}{i\om_{\nu}-\om'-\om_0}\right)\,,\nonumber 
\end{eqnarray}
with the fermionic Matsubara frequencies $\omega_\nu=(2\nu+1)\pi/\beta$ and the Fermi function $n_F=(\me^{\beta\omega}+1)^{-1}$.
Analytical continuation $\mi\omega_\nu\to\bar{\omega}=\om+\mi\delta$
in the upper complex half-plane then gives the retarded Green function
\begin{equation}
\label{rdd_gf}
G_{dd}^R(\bar{\omega})=\frac{1}{\bar{\omega}-(\tilde\Delta-\mu)-\Sigma_{dd}(\bar{\omega})}
\end{equation}
and the related spectral function
\begin{equation}
\label{rdd_sf}
A_{dd}(\om)=-\frac{1}{\pi}\Im G_{dd}^R(\om+\mi0^+)\,.
\end{equation}
The r.h.s. of equation~(\ref{rdd_sf}) is determined by the real and imaginary
parts of $\Sigma_{dd}(\om+\mi 0^+)$, which we obtain  from~(\ref{sigmatwocomplex}) employing $\frac{1}{x+\mi 0^+}={\cal P}\frac{1}{x}-\mi \pi \delta(x)$.  
Moreover, we transform the $k$-summation into an integration over the 
band energy $\xi$ of the leads, $\xi \in [-W,W]$, using the lead density of states $\varrho(\xi)$. Assuming 
further the right and the left leads to be identical the summation over
$a$ gives simply a factor 2. In the end, we work in the limit $T\to 0$, 
when the Fermi function becomes the Heaviside function, $n_F(\omega)=\Theta(-\om)$,
and obtain    
\begin{eqnarray}\label{im_sigma}
&&\Im \Sigma_{dd}(\om) =
-2\pi\,t_d^2\,\me^{-\tilde{g}^2}\varrho(\om+\mu)\int_{-W}^{W}\mathrm{d}\xi\, 
\delta[\om-(\xi-\mu)] \\
&&\quad- 2 \pi\,t_d^2 \,\me^{-\tilde g^2}\sum_{s\geq 1}\frac{(\tilde g^2)^{s}}{s!}
\Big\{ \varrho(\om+\mu+s\om_{0})
	\int_{-W}^{\mu}\mathrm{d}\xi \,
\delta [\om -(\xi-\mu) +s\om_{0}]\nonumber \\
	&&\quad+ \varrho(\om+\mu-s\om_{0})   
\int_{\mu}^{W}\mathrm{d}\xi\, \delta 
[\om-(\xi-\mu)-s\om_{0}]\Big\}\nonumber\\[0.2cm]
&&\quad -\pi\left[(1-\gamma)g\om_0\right]^2\left\{A_{dd}(\om+\om_0)\Theta(-\om-\om_0)+ A_{dd}(\om-\om_0)\Theta(\om-\om_0)\right\} \nonumber\,,
\end{eqnarray}
\begin{eqnarray}\label{re_sigma}
&&\Re \Sigma_{dd}(\om)=2\,t_d^2\,\me^{-\tilde g^2} \mathcal{P} \int_{-W}^{W}
\mathrm{d}\xi\, \varrho(\xi) \frac{1}{\om-(\xi-\mu)}\\
&&\quad 2\,t_d^2\,\me^{-\tilde g^2}\sum_{s\geq 1}\frac{(\tilde g^2)^{s}}{s!}\mathcal{P} \left \{ \int_{-W}^{\mu}\mathrm{d}\xi\, \varrho(\xi)  \frac{1}{\om-(\xi-\mu)+s\om_{0}}\right .\nonumber \\
	&&\hspace*{4cm}+ \left . \int_{\mu}^{W}\mathrm{d}\xi \,\varrho(\xi) \frac{1}{\om-(\xi-\mu)-s\om_{0}} \right\}\nonumber\\
&&\quad+
\left[(1-\gamma)g\om_0\right ]^2  \left \{ \mathcal{P} \int_{-\infty}^{0} \mathrm{d}\om'  \frac{A_{dd}(\om')}{\om+\om_0-\om'}+ \mathcal{P}\int_{0}^{\infty} \mathrm{d}\om'  \frac{A_{dd}(\om')}{\om-\om_0-\om'}\right\}\,.\nonumber
\end{eqnarray}
Note that in accordance with our second order approach, the spectral functions 
$A_{dd}(\om)$ occurring in~(\ref{im_sigma}), (\ref{re_sigma}) have to be
determined by equations~(\ref{rdd_gf}), (\ref{rdd_sf}), using
$\Sigma_{dd}^{(1)}(\om+\mi 0^+)$ for the self-energy.

For $g=0$, we are faced with the well-known problem of electron localisation
at an impurity. Then, if a solution $\tilde{\om}$ of  
\begin{equation}
\label{polsolution}
\tilde{\om}= \Delta+\Re \Sigma_{dd}(\tilde{\om}-\mu)
\end{equation}
exists outside the interval $[-W,W]$, the spectral function $A_{dd}(\om)$
exhibits a single-peak structure,
\begin{equation}\label{a_dd_dot}
A_{dd}(\om)=z(\tilde{\om})\delta[\om-(\tilde{\om}-\mu)]\,,
\end{equation}
reflecting electron localisation at the quantum dot. Here, 
\begin{equation}\label{z-factor}
z(\tilde{\om})^{-1}=\left | 1 + 2 \,t_d^2 \, \mathcal{P}\int_{-W}^{W}\mathrm{d}\xi \, \varrho(\xi) \frac{1}{(\tilde{\om}-\xi)^2}\right |\,. 
\end{equation}
According to the first term on the r.h.s. of~(\ref{im_sigma}),
we find $\Im \Sigma_{dd}(\om)\neq 0$ for $\om \in [-W-\mu,W-\mu]$,
leading to an incoherent single-particle spectrum in this interval.

\begin{figure}[b]
\hspace*{0.6cm}\includegraphics[width=0.9\textwidth]{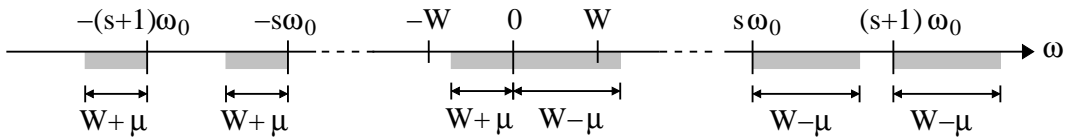}
\caption{Shaded regions indicate the intervals of non-zero contributions
to $\Im \Sigma_{dd}(\om)$ from the processes determined by the 
first two terms on the r.h.s. of equation~(\ref{im_sigma}). 
Note that all intervals overlap 
if $\om_0 < W-|\mu|$. 
Shown is the case $\om_0>W$, $\mu<0$. 
}
\label{Fig_intervals}
\end{figure}
For non-zero coupling between the electron and the local vibrational mode
at the dot, we have  $\Im \Sigma_{dd}(\om)\neq 0$ everywhere 
provided that $\om_0 < W-|\mu|$. 
Then, if $g\neq 0$, the spectral function 
$A_{dd}(\om)$ exhibits no coherent contribution, and is given,
for all $\om$ and $\gamma \in [0,1]$, by a purely
incoherent spectrum 
\begin{equation}\label{inc_spectrum}
A_{dd}(\om)=-\frac{1}{\pi}\frac{\Im\Sigma_{dd}(\om)}{\left [\om-(\tilde \Delta-\mu)-\Re\Sigma_{dd}(\om)\right ]^2+\left [\Im\Sigma_{dd}(\om)\right ]^2}\;\;.
\end{equation}
On the other hand, if the latter condition is not fulfilled
because $-\om_0<-W-\mu$ or $W-\mu<\om_0$, then the contributions to  
$\Im\Sigma_{dd}(\om)$ from the phonon processes vanish 
in certain $\om$-intervals 
(see figure~\ref{Fig_intervals}). In particular, this happens
for $\om<0$, $\om \in [-(s+1)\om_0,-s\om_0-W-\mu]$ and for 
$\om>0$, $\om \in [W-\mu+s\om_0, (s+1)\om_0]$ with $s\geq 0$, respectively.
Poles in the Green function occur if equation~(\ref{polsolution}) has a real-valued solution in these intervals. 
A small width of these peaks can arise from the third
term on the r.h.s. of equation~(\ref{im_sigma}) if $\gamma$ differs 
appreciable from unity.

\subsection{Determination of $\gamma$}
In order to fix the variational parameter $\gamma$ self-consistently, 
we minimise the ground-state expectation value $E=\langle \tilde H \rangle $ with respect to $\gamma$. We factorise the statistical averages with respect to phonon and polaron variables, i.e. $\langle C_{d}d^{\dag} d \rangle \approx \langle C_{d} \rangle \langle d^{\dag} d \rangle $ and $\langle C^{\phantom{\nag}}_{t}d^{\dag}c_{ka}^{\nag} \rangle \approx \langle C^{\phantom{\nag}}_{t} \rangle \langle d^{\dag}c_{ka}^{\nag}\rangle $ and obtain
\begin{eqnarray}\label{hamiltonian_mean_value}
E & = &  \sum_{k,a} (E_{ka} -\mu) 
\langle c_{ka} ^\dag c_{ka}^{\nag} \rangle - 
\sum_{k,a} \left ( \langle C_{t}\rangle\langle  d^\dag c_{ka}^{\nag} \rangle + \langle C_{t}^\dag\rangle\langle  c_{ka}^\dag d \rangle \right )   \nonumber\\
& & \quad+(\tilde \Delta-\mu) \langle d^\dag d \rangle - \langle C_d\rangle\la d^\dag d \ra + \omega_0 \langle b^\dag b \rangle \;.
\end{eqnarray}
For the transformed Hamiltonian,
$\langle b^{\dagger}b^{}\rangle$ as well as 
$\langle C_d \rangle$ is zero for $T\to 0$ in our second order approach.
The remaining expectation values
can be expressed as 
\begin{eqnarray}
\langle d^{\dag}d^{} \rangle &=& \int_{-\infty}^\infty \mathrm{d} \omega' A_{dd}(\om')n_F(\om')\,,\label{ddagd}\\
\langle d^\dag c_{ka} \rangle & = & G_{cd;\,ka}(\tau_1,\tau_2) \Big |_{\tau_1^{\phantom{}} \to \tau_2^-}
  = -\frac{1}{\pi} \int_{-\infty}^\infty \mathrm{d} \omega'\, n_F(\om')\, \Im  G_{cd;\,ka}(\om') \label{ddagc}\,, \\
\langle c_{ka}^\dag c_{ka}^{\nag} \rangle & = & G_{cc;\,ka}(\tau_1,\tau_2) \Big |_{\tau_1^{\phantom{}} \to \tau_2^-}
  = -\frac{1}{\pi} \int_{-\infty}^\infty \mathrm{d} \omega'\, n_F(\om')\,  \Im  G_{cc;\,ka}(\om')  \,.
\end{eqnarray}
%
%
Based on the equations of motion (equation (\ref{g0cc-1}) for $G_{cd;ka}$ and an
analogous one for $G_{cc;ka}$), the Green functions are determined to first
order as
\begin{eqnarray}
G_{cd;ka}(z)&\approx& - G_{cc;ka}^{(0)}(z)\langle C^{\dag}_{t} \rangle G_{dd}(z)\;,\label{appr_gcd}\\
G_{cc;ka}(z)&\approx &G_{cc;ka}^{(0)}(z)+ \langle C^{\dag}_{t} \rangle \langle C^{\nag}_{t} \rangle \big [ G_{cc;ka}^{(0)}(z) \big ]^{2} G_{dd}(z)\;,\label{gccapprox}
\end{eqnarray}
which is consistent with the factorisation of bosonic and polaronic averages in (\ref{hamiltonian_mean_value}).
Via $G_{dd}$, the expressions (\ref{appr_gcd}) and (\ref{gccapprox}) depend on the spectral function $A_{dd}$. We
identify the variational parameter $\gamma$ from the minimum of $E(\gamma)$.
\subsection{Two-particle properties: kinetic coefficient}\label{sec:kincoeff}
In this section we calculate the current of the system~(\ref{ho1})
caused by a small potential difference between the leads.
To this end we deduce the expression for the kinetic coefficient in terms 
of the quantum-dot spectral function, consistently with 
the approximations assumed in the derivation of $A_{dd}(\om)$. 
In accordance with general linear response theory~\cite{Zu71},
the external perturbation $H_t=-\alpha F(t)$ coupled to the system variable
$\alpha$ induces the change $\langle \dot{\alpha} \rangle$ of $\langle\alpha\rangle$, whose Fourier transform $\langle \dot{\alpha} \rangle_{\omega}$ is related to the Fourier
component $F(\om)$ via the kinetic coefficient $L(\omega)$ as  
\begin{equation}
\la \dot \alpha \ra_{\omega}  =  L(\omega)F(\omega)\,.
\end{equation}
In particular, linear response theory gives for $L=L(\om \to 0)$
the expression
\begin{equation}\label{L_coefficient}
L  =  \lim_{\omega\to 0} \left[ - \frac{1}{\omega} \Im 
\langle\langle\dot \alpha | \dot \alpha\rangle\rangle_{\omega} \right]\,,
\end{equation} 
where the symbol $\langle\langle\cdot | \cdot\rangle\rangle_\om$ denotes the retarded (commutator)
Green function in frequency representation. In our case, 
the quantity of interest is the current between
the lead $a$ and the dot. By the continuity equation, the operator for this
current is
\begin{equation}\label{conti}
J_a = -e\dot N_a\,,
\end{equation}
where $N_a  =  \sum_k c^\dag_{ka}c_{ka}^{}$ and 
\begin{equation}
\dot N_a = \mi [H,N_a] = -\mi\frac{1}{\sqrt{N}}\,t_d \sum_k  \left ( \tilde d ^\dag c_{ka} - c^\dag_{ka} \tilde d \right ) \label {deriv}
\end{equation}
with the electron $\tilde d$ operators defined in~(\ref{traop}). 
With $\alpha=N_a$ in the above general linear response formula,
we express the kinetic coefficient, obtained from variation of the chemical potential $\mu_a=\mu+\Delta\mu_a$ in the first term of the Hamiltonian,
through the retarded Green function $\langle\langle \dot{N}_a |\dot{N}_a\rangle\rangle$.

To obtain this Green function,
let us consider the corresponding Matsubara Green function
\begin{equation}
G_J(\tau_1,\tau_1')  =  - \la \mathcal{T}_\tau \dot N_a (\tau_1) \dot N _a (\tau_1') \ra 
\end{equation}
which is related to the two-particle Green function by  
\begin{equation}\label{gjlimit}
G_J(\tau_1,\tau_1')  =  \bar G (\tau_2,\tau_2',\tau_1,\tau_1') \Big |_ { \tau_2^{\phantom{}}  =   \tau_1^-,  \tau_2'  =  \tau_1'^- }  \,, 
\end{equation}
where
\begin{eqnarray}
\bar G (\tau_2,\tau_2';\tau_1,\tau_1')  & = & \frac{1}{N}\,t_d^2 \sum_{k,k'} 
	\left [ \la \mathcal{T}_\tau c_{ka}(\tau_2) c_{k'a}(\tau_2') \tilde d ^\dag (\tau_1') \tilde d ^\dag (\tau_1) \ra \right .  \label{Gbar}  \nonumber \\
& &\qquad -	\la \mathcal{T}_\tau c_{ka}(\tau_2) \tilde d(\tau_2') c_{k'a} ^\dag (\tau_1') \tilde d ^\dag (\tau_1) \ra   \nonumber\\
& & \qquad -	 \la \mathcal{T}_\tau \tilde d(\tau_2) c_{k'a} (\tau_2') \tilde d ^\dag  (\tau_1') c_{ka} ^\dag (\tau_1) \ra  \nonumber\\
& &\qquad  +	\left.  \la \mathcal{T}_\tau \tilde d(\tau_2) \tilde d (\tau_2') c_{k'a} ^\dag  (\tau_1') c_{ka} ^\dag (\tau_1) \ra  \right ]
\end{eqnarray}
The mean values of all the time-ordered products on the r.h.s. of~(\ref{Gbar})
are basically two-particle Green functions $G(2,2^\prime;1,1^\prime)$~\cite{Ri81}, which may be approximated by one-particle Green functions according to
\begin{equation}\label{HF}
G(2,2';1,1')  \approx  G(2,1)G(2',1') - G(2,1')G(2',1)\,,
\end{equation}
if vertex corrections due to phonon-mediated electron-electron scattering are neglected.
In this way, the terms on the r.h.s. of~(\ref{Gbar}) turn out to be 
proportional to products of $G_{c\tilde{d}}$,  $G_{\tilde{d}c}$,  
$G_{cc}^{}$, and $G_{\tilde{d}\tilde{d}}$. Substituting $\tilde{d}$ 
from~(\ref{traop}), averaging over the oscillator variables and inserting
the expression~(\ref{appr_gcd}) for $G_{cd}$ and  $G_{dc}^{}=G_{cd}^*$,  
it becomes obvious that the terms containing the latter ``mixed''
Green functions are of higher order in $|\langle C_{t}\rangle|$.
Hence we get as the leading order result
\begin{eqnarray}\label{gtilde}
\bar G (\tau_2,\tau_2',\tau_1,\tau_1')  & = & \frac{1}{N} \,t_d^2\sum_{k} 
 \left [  G_{cc;\,ka}^{(0)} (\tau_2,\tau_1') \tilde G_{dd}(\tau_2',\tau_1)\right.\nonumber\\&&\hspace*{3cm}\left. + \tilde G_{dd}(\tau_2,\tau_1')  G_{cc;\,ka}^{(0)} (\tau_2',\tau_1) \right ]\,.
\end{eqnarray}
Here we have used the unperturbed Green function $G^{(0)}_{cc}$ (see equation~(\ref{Green_unpert_lead}))
for the electrons in the leads and introduced the notation 
$\tilde{G}_{dd}=G_{\tilde{d}\tilde{d}}$.

Inserting~(\ref{gtilde}) into~(\ref{gjlimit}) and performing
a Fourier transformation, the latter equation becomes 
\begin{eqnarray}\label{GJproduct}
G_J(\mathrm{i} \omega_n) & = & \frac{1}{N}\, t_d^2 \sum_{k} 
\frac{1}{\beta}\sum_{\om_\nu}  \left [  G_{cc;\,ka}^{(0)} (\mathrm{i} \omega_\nu + \mathrm{i} \omega_n) \tilde G_{dd}(\mathrm{i} \omega_\nu)\right.\nonumber\\
&&\hspace*{3.5cm}\left. + 
   G_{cc;\,ka}^{(0)} (\mathrm{i} \omega_\nu) \tilde 
G_{dd}(\mathrm{i} \omega_\nu+\mathrm{i} \omega_n)  \right ]\,,
\end{eqnarray}
with bosonic Matsubara frequencies $\omega_n$.  We now express $\tilde G_{dd}$ by the
electronic spectral function $\tilde A_{dd}(\om)$
\begin{equation}\label{ttildeg}
\tilde G_{dd}(\mathrm{i} \omega_\nu)=\int_{-\infty}^\infty \mathrm{d}\om'
\frac{\tilde{A}_{dd}(\om')}{\mathrm{i} \omega_\nu-\om'}\,,
\end{equation}
make use of
\begin{equation}\label{gcca}
G_{cc;\,ka}^{(0)}(\mathrm{i} \omega_\nu)=\frac{1}{\mathrm{i} \omega_\nu-(E_{ka}-\mu)}\,,
\end{equation}
perform the Matsubara summation over the fermionic frequencies $\omega_\nu$, and obtain 
\begin{eqnarray}\label{gjmatsu}
G_J(\mathrm{i} \omega_n) & = & \frac{1}{N} \,t_d^2\sum_{k}
 \left \{   \int_{-\infty}^{+\infty}\mathrm{d}\omega'   \, \tilde A_{dd} (\omega') 
	\frac{n_F(\omega') - n_F(E_{ka}-\mu)}{\mathrm{i}\omega_n + \omega'  -(E_{ka}-\mu)} \right. \nonumber\\
& & + \left.  \int_{-\infty}^{+\infty}\mathrm{d}\omega'   \,  \tilde A_{dd} (\omega') 
	\frac{n_F(E_{ka}-\mu) - n_F(\omega')}{\mathrm{i}\omega_n - \omega' + (E_{ka}-\mu)}   \right \}\,.
\end{eqnarray}
The analytical continuation of (\ref{gjmatsu}),  
$\mi \om_n \to \bar{\omega}=\om+\mi \delta$, 
gives the retarded Green function
\begin{equation}
 \las \las {\dot N_a} | {\dot N_a} \ras \ras_\omega  =  
G_J(\omega+\mathrm{i} 0^+)\,,
\end{equation}
 leading to
\begin{eqnarray}
\hspace{-1.5cm}\Im \langle\langle\dot N_a | \dot N_a\rangle\rangle_\omega\!\!&=&\!\!- 
\frac{\pi}{N} \,t_d^2 \sum_{k} 
\left \{  \tilde A_{dd} (E_{ka}-\mu - \omega)
\left [ n_F(E_{ka}-\mu - \omega) - n_F(E_{ka}-\mu) \right ]\right.\nonumber\\ 
&&\left.+ 
\tilde A_{dd} (E_{ka}-\mu + \omega) 
\left [ n_F(E_{ka}-\mu) - n_F(E_{ka}-\mu + \omega) \right ]  \right \}\,.
\end{eqnarray}
Assuming identical leads $a=l,r$, this result is of course independent of $a$.
Finally, according to the definition~(\ref{L_coefficient}), 
we have to perform the limit $\omega\to 0$:
\begin{eqnarray}
L & = & \lim_{\omega \to 0} \left [ - \frac{1}{\omega} \Im \langle\langle\dot 
N_a | \dot N_a\rangle\rangle_{\omega} \right ] = \frac{2\pi}{N}\,t_d^2 \sum_{k} \tilde A_{dd} 
(E_{k}-\mu) \left ( - n_F'(E_{k}-\mu) \right ) \nonumber  \\\
& = & 2 \pi\,t_d^2\int_{-W}^{W} \mathrm{d}\xi \,  \varrho(\xi) \tilde A_{dd} (\xi-\mu)\, 
\left ( - n_F'(\xi-\mu) \right )\;.
\end{eqnarray}
Then, for $T\to 0$, $(- n_F'(\xi-\mu))=\delta[\xi-\mu]$, so that
\begin{equation}\label{LfromA}
L  =  2  \pi\,  t_d^2 \,\varrho(\mu) \tilde A_{dd} (0)\,. 
\end{equation}
The general relation between the electronic spectral function 
$\tilde{A}_{dd}(\om)$ needed here and the polaronic spectral function 
$A_{dd}(\om)$ determined in the preceding section, 
was derived in Ref.~\cite{LHF06}. Using equation 
(40) of this work, for $\om=0$ and $T\to 0$, it simply follows that    
\begin{equation}\label{tildeAfromA}
\tilde A_{dd}(0)  =  \me^{-\tilde g^2} A_{dd}(0)\,.
\end{equation}
Consequently, the kinetic coefficient is determined by the value
of the polaronic spectral function $A_{dd}(\om)$ at the lead Fermi level, multiplied with the renormalisation factor $\me^{-\tilde g^2}$. 
In particular, changing $g$ or $\Delta$ leads to a shift of $A_{dd}(\om)$
with respect to the lead Fermi level and therefore changes the value
of $A_{dd}(0)$, as will be shown by the numerical calculation in the 
following section.
\section{Numerical results}\label{sec:results}
While the derivation of the equations in section~\ref{sec:theory} is completely general, we consider in the following the case of a single quantum dot between
semi-infinite 1D leads, as sketched in figure~\ref{hu_model}.
Accordingly, the density of states of the leads is given by

\begin{equation}\label{dos}
\varrho(\xi)=\frac{2}{\pi W} \sqrt{1-(\xi/W)^2}\;\Theta(1-(\xi/W)^2)
\end{equation}
with the half-bandwidth $W=2t$.
We fix $t=1$ from here on.

The numerical computation of the Green function $G_{dd}(\bar{\omega})$, equation~(\ref{rdd_gf}), and the corresponding spectral function $A_{dd}(\omega)$, equations~(\ref{rdd_sf}),~(\ref{inc_spectrum}), is performed by evaluation of equation~(\ref{sigmatwocomplex}) for energies $\bar{\omega} = \omega +  \mi \delta$ slightly above the real axis.
A small choice of $\delta>0$ avoids problems arising from the simultaneous treatment of poles and incoherent parts in the Green function.
For our computations, we used $\delta \simeq 10^{-3}$.
Alternatively, one might directly evaluate equations~(\ref{im_sigma}),~(\ref{re_sigma}) which are given for real $\omega$, but numerical inaccuracies in the calculation of the principal value integrals tend to degrade the computation.
From the spectral function, we obtain the kinetic coefficient $L$ using equations~(\ref{LfromA}),~(\ref{tildeAfromA}).

It is noteworthy that, restricting ourselves to second order perturbation,
the approximations to the Green functions preserve important sum-rules
for the spectral functions, especially for the integrated weight~\cite{Ko09}.
This remains true for finite $\delta>0$ in the numerical computation
(while the numerically unfavourable evaluation of equations~(\ref{im_sigma}),~(\ref{re_sigma}) led to the artificial drop of total spectral weight reported in Ref.~\cite{LHF06}).

In the discussion of the numerical results we start with important 
limiting cases.

\subsection{Non-interacting case}\label{sec:NoInter}

\begin{figure}
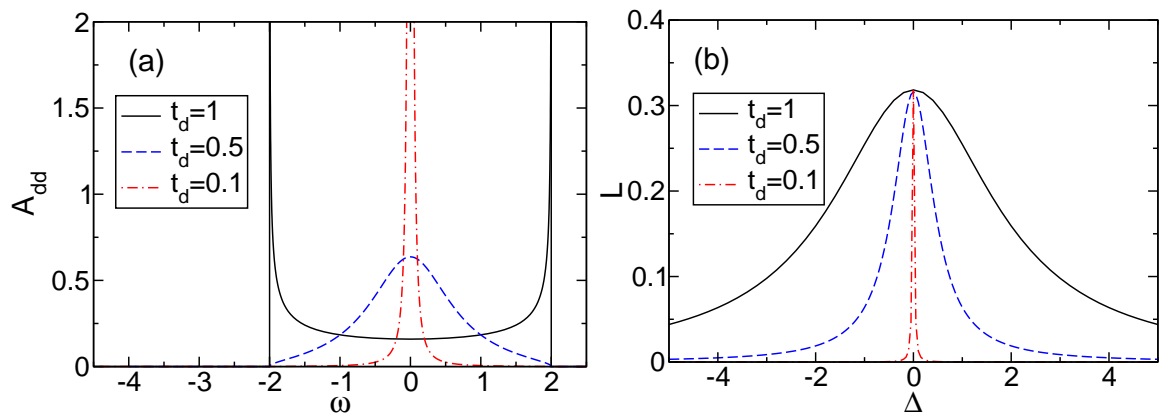

  \begin{center}
\includegraphics[width=0.49\textwidth]{fig2a}
\includegraphics[width=0.475\textwidth]{fig2b}
  \end{center}
\caption{(Colour online) All results for $\varepsilon_p=0$ and $\mu=0$.
Left panel: Dot spectral function $A_{dd}(\omega)$ for $\Delta=0$ and several $t_d$.
Right panel: Kinetic coefficient $L$ as a function of $\Delta$ for several $t_d=1$.}
\label{Fig2}
\end{figure}

For vanishing EP coupling ($\varepsilon_p=0$) the problem reduces to that of an
impurity in a 1D chain. The spectral function is then obtained exactly
by our calculation (left panel in figure~\ref{Fig2}). For $t_d=1$, the
spectral function of the translationally invariant 1D chain is
recovered. For smaller $t_d$, the spectral function develops a pronounced
maximum at $\omega=0$, which evolves into a $\delta$-peak in the 
limit $t_d \to 0$.
 
For the kinetic coefficient, we observe in figure~\ref{Fig2} (right panel) 
two effects that will be important later in our discussion of the interacting case.
First, for dot energy $\Delta \ne 0$ scattering off the dot impurity leads to  reduction of $L$ compared to the case $\Delta=0$ with minimal scattering. 
Consequently, $L$ is maximal for $\Delta=0$ and shrinks monotonically with growing $|\Delta|$.
Second, a reduction of $t_d$ leads to a reduction of $L$.
Moreover the variation with $\Delta$ becomes more pronounced as electrons
become more susceptible to scattering off the dot.
In equations~(\ref{im_sigma}),~(\ref{re_sigma}) we see
that for smaller hybridisation $t_d^2 \,\varrho(\xi)$ 
the broadening of dot levels due to coupling to the continuum of lead
states is reduced.
In the limit $t_d \to 0$, the function $L=L(\Delta)$ becomes a $\delta$-function at $\Delta=0$ with weight $\propto t_d^2$.
Note that $L$ is independent of $t_d$ for
$\Delta=0$, which is however a peculiarity of the non-interacting case
without damping of states close to the Fermi energy. 
%


\subsection{Small phonon frequency}

In the following, we first discuss the results of our approach 
in the limits of small and large (next subsection) phonon frequencies.
For the moment, we fix $t_d = t = 1$ and $\mu=0$ which
corresponds to the half-filled band case for a translational 
invariant system with $\varepsilon_p=0$ and $\Delta=0$.  

For small phonon frequency $\omega_0/t = 0.1$ (adiabatic regime), 
when the phononic timescale is much slower than the electronic 
timescale, we expect significant deviations from the behaviour
described by the standard Lang-Firsov approach consisting of
a complete Lang-Firsov transformation and a subsequent average 
over the transformed phonon vacuum. Our approach is able to account 
for these deviations by the variational parameter $\gamma$. 
The deviation of $\gamma$ from unity is some measure of 
both adiabatic and weak-coupling corrections.

\subsubsection{Repulsive dot}

\begin{figure}[hb]
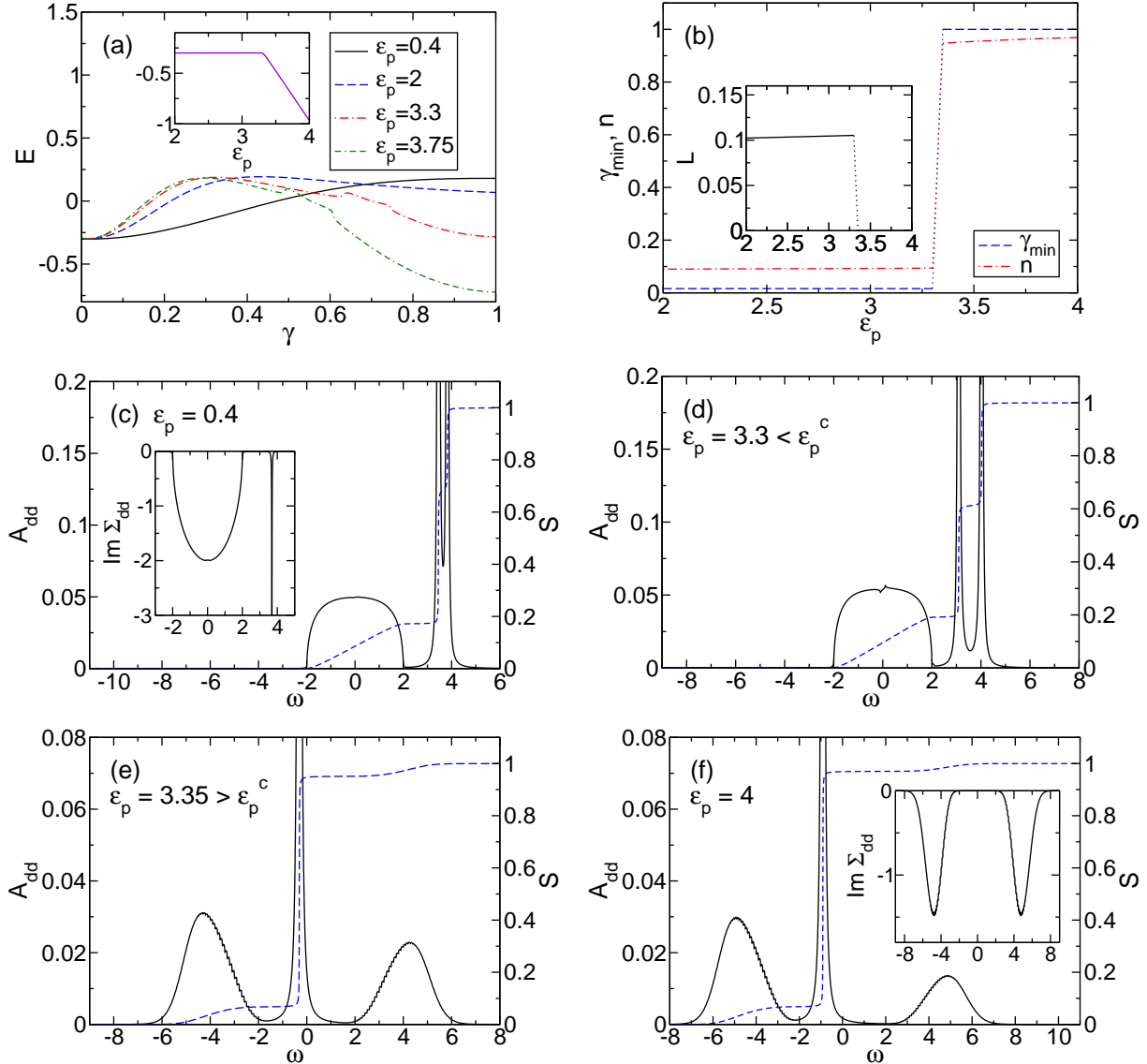

    	\hspace*{0.1cm}\includegraphics[width=0.44\textwidth]{fig3a}\hspace{1.25cm}
    	\includegraphics[width=0.44\textwidth]{fig3b}\\[0.25cm]
    	\includegraphics[width=0.5\textwidth]{fig3c}\hspace{0.3cm}
    	\includegraphics[width=0.5\textwidth]{fig3d}\\[0.25cm]
    	\includegraphics[width=0.5\textwidth]{fig3e}\hspace{0.3cm}
    	\includegraphics[width=0.5\textwidth]{fig3f}
  \caption{(Colour online) All results for $\Delta=3$, $t=t_d=1$, 
$\mu=0$, and $\omega_0=0.1$. 
Upper panels: Total energy $E$ (a) as a function of $\gamma$, 
for different $\varepsilon_p$ (inset: Minimum of $E$ as a function 
of $\varepsilon_p$) and optimal parameter $\gamma_{min}$, kinetic 
coefficient $L$ and particle density $n$ at the dot (b) as a function of $\varepsilon_p$. 
Lower panels: Dot spectral function $A_{dd}(\omega)$ and 
integrated spectral weight $S(\om)$ for  different 
$\varepsilon_p$ (c)-(f). The insets in panels (c) and (f) 
show the imaginary part of the second-order self-energy 
$\Sigma_{dd}(\omega)$.}
  \label{FigAdiab1}
\end{figure}

In figure~\ref{FigAdiab1} we show results for the case $\omega_0 = 0.1$,
$\mu=0$, and a repulsive dot $\Delta = 3$.
Since the dot is repulsive, the particle density  at the dot is small, 
$n\simeq 0.1$, for small EP coupling $\varepsilon_p$ [see panel (b)].
At a critical coupling 
$\varepsilon_p^c \simeq 3.325$, EP interaction at the dot overcomes the 
repulsive potential, and a transition takes place to a situation with large $n$. This transition is accompanied by a jump of the variational parameter $\gamma_{min}$ from a small value ($ < 0.02$) to $1$.
This jump can be traced back to the behaviour of the total energy $E$ as a function of $\gamma$: If $\varepsilon_p$ increases, $E(\gamma)$ develops 
two local minima [see panel (a)]. At $\varepsilon_p=\varepsilon_p^c$, the 
minimum at $\gamma=1$ becomes the new global minimum.
Evidently, the sudden change of $\gamma$ reflects the formation of a 
strongly localised polaron at the quantum dot. 
Thereby the lead-dot transfer is almost completely suppressed
and, in accordance with this picture, the kinetic coefficient 
($\propto \exp\{-g^2\}$) drops to zero at the transition point. 
This suppression of transport
is well described by the complete Lang-Firsov transformation 
(although this  basically is a non-adiabatic approach), mainly because
we enter an extreme strong-coupling situation ($g^c=5.77$). 
Note that the observation of an extremely sharp polaron 
transition in the adiabatic regime  for repulsive quantum dots 
is in accordance with recent exact diagonalisation results~\cite{FWLB08,AF08b}.    

We next analyse the dot spectral function $A_{dd}$
[see figure~\ref{FigAdiab1} (c)-(f)]. 
For $\varepsilon_p < \varepsilon_p^c$,  we have 
$\tilde{g}\leq\gamma_{min} g^c<0.1154$ and, calculating $\Sigma_{dd}^{(1)}$ 
within our second order scheme, the first term in equation 
(\ref{sigmatwocomplex}) is basically proportional to the 
semi-elliptical density of states of the leads, while the 
second term (``phononic contribution'') is insignificant. 
Then the resulting spectral function $A_{dd}^{(1)}$,   
which has to be put into the third term of (\ref{sigmatwocomplex}), 
describes a continuum of states, roughly in between -2 and 2,
and a localised dot state at $\om\simeq 3.5$ (in accordance   
with the result obtained for the Fano-Anderson model).
Since the prefactor of the third term, $(1-\gamma_{min})g\om_0\lesssim 0.57$
for $g\leq g^c$, 
is rather large this term gives a significant contribution to the second
order self-energy $\Sigma_{dd}$. Thereby the localised peak 
in $A_{dd}^{(1)}$ becomes evident in  $\Sigma_{dd}$
[see inset of figure~\ref{FigAdiab1}~(c)]. As a result
the second order spectral function, $A_{dd}$, exhibits 
two sharp peak structures  
(localised states) above the continuum of states around $\omega=0$. 
If the EP coupling increases these peaks become more and more separated.
In order to analyse the spectral weight of the different signatures
in $A_{dd}$, we have calculated the integrated spectral function
\begin{equation}\label{int_weight}
S(\om)=\int_{-\infty}^{\om}\, \mathrm{d}\om' A_{dd}(\om')\,.
\end{equation}
Figures~\ref{FigAdiab1}~(c) and~(d) show that 
for $\varepsilon_p < \varepsilon_p^c$ the spectral weight 
mainly rests in the localised peak structures above the wide band.
Hence the spectral weight of the current--carrying states at the Fermi 
energy $\mu=0$ is reduced, and the kinetic coefficient
$L\propto A_{dd}(0)$ is substantially lowered compared to 
the case $\Delta=0$. At $\varepsilon_p = \varepsilon_p^c$, 
$\gamma_{min}$ jumps to 1, and the strong renormalisation arising 
from the complete Lang-Firsov-transformation results in a pronounced 
peak at negative energy at about $\Delta-\varepsilon_p$, which now, however, 
is the signature of a quasi-localised polaronic dot state.   
The polaronic quasiparticle peak is accompanied by two side-bands 
 (roughly of width $2W$) shifted by $\pm \varepsilon_p$, 
which arise from the Poissonian distribution of phonons at the dot, 
with maximum at $g^2 = \varepsilon_p/\omega_0$ phonons. 
States in these band are strongly damped due to the significant phononic admixture, as is evident in the imaginary part of the self-energy [see inset
figure~\ref{FigAdiab1}~(f)].

\subsubsection{The case $\Delta=0$}
For $\Delta < 0$ the quantum dot is attractive.
For $\Delta =0$ and $\varepsilon_p=0$ we have of course a 
translational invariant 1D system, where $\mu=0$ corresponds to the
half-filled band case, i.e., $n=0.5$. Such a ``neutral'' quantum dot
becomes attractive for arbitrarily weak EP interaction. This is because the
``effective'' dot level is given by $\tilde\Delta=\Delta-\varepsilon_p
\gamma_{min}(2-\gamma_{min})$. 
\begin{figure}[b]
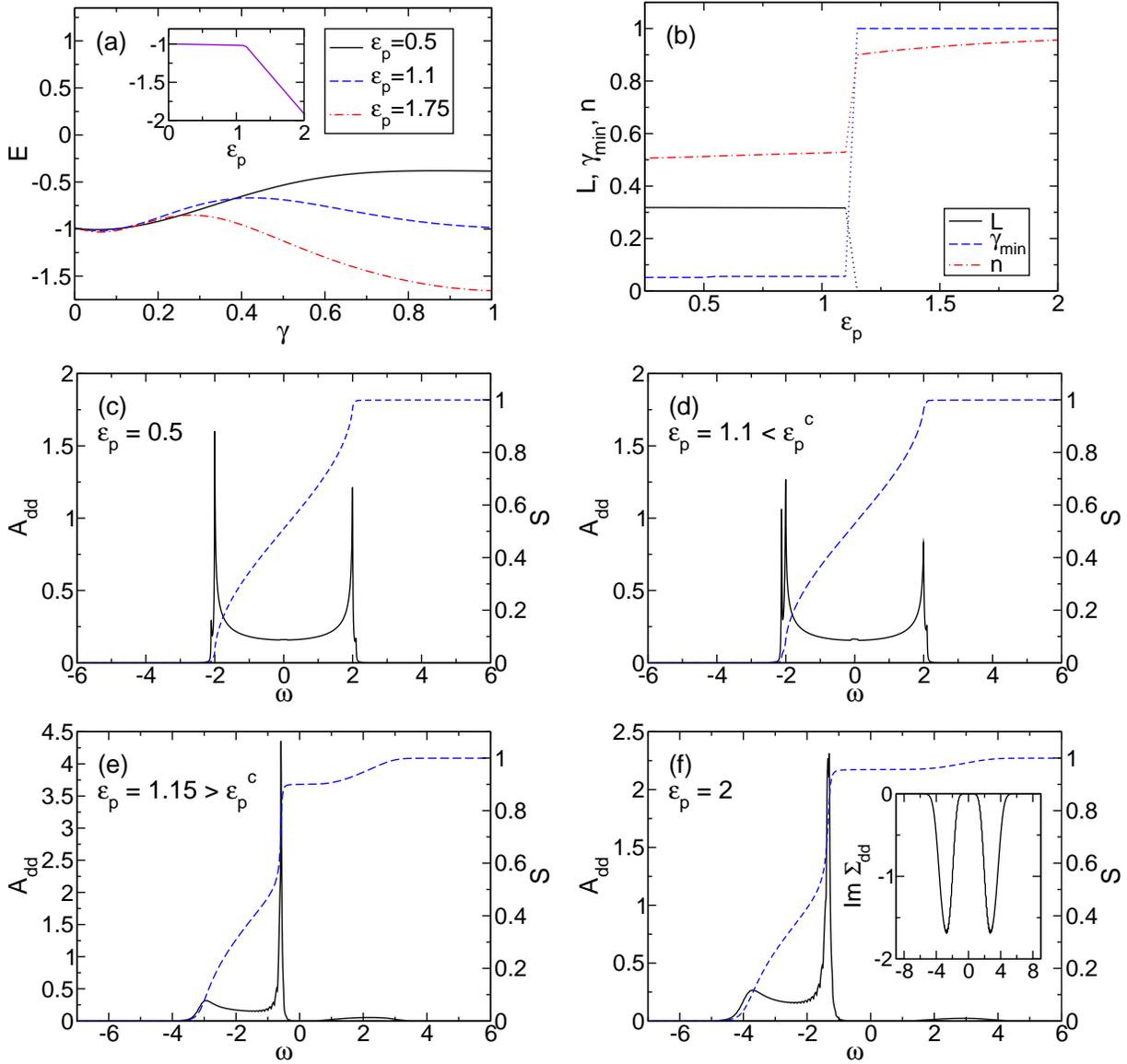

    	\includegraphics[width=0.45\textwidth]{fig4a}\hspace{1.0cm}
    	\includegraphics[width=0.45\textwidth]{fig4b}\\[0.25cm]
    	\includegraphics[width=0.5\textwidth]{fig4c}\hspace{0.3cm}
    	\includegraphics[width=0.5\textwidth]{fig4d}\\[0.25cm]
    	\includegraphics[width=0.5\textwidth]{fig4e}\hspace{0.3cm}
    	\includegraphics[width=0.5\textwidth]{fig4f}
  \caption{(Colour online) All results for $\Delta=0$, $t=t_d=1$, 
$\mu=0$, and $\omega_0=0.1$.
Upper panels: Total energy $E$ (a) as a function of $\gamma$, for different $\varepsilon_p$ (inset: Minimum of $E$ as a function of $\varepsilon_p$) and  
optimal parameter $\gamma_{min}$, kinetic coefficient $L$ and particle density $n$ at the dot (b) 
as a function of $\varepsilon_p$. 
Lower panels: Dot spectral function $A_{dd}(\omega)$ and 
integrated spectral weight $S(\om)$ for different $\varepsilon_p$ (c)-(f). 
The inset in panel (f) gives the imaginary part of the dot self-energy 
$\Sigma_{dd}(\omega)$.
}
  \label{FigAdiab2}
\end{figure}

Consequently, in figure~\ref{FigAdiab2}~(b) the particle density at the 
dot is larger than $0.5$ for all $\varepsilon_p$, and the dot 
spectral function has no pole at positive energies. At small EP coupling
the spectral function is similar to that of of a 1D tight-binding model.
The weak EP interaction causes the spiky signatures separated by $\om_0$
from the upper and lower band edges [see figure~\ref{FigAdiab2}~(c)]. 
In contrast to a repulsive dot, for which our methods correctly describes 
the transition from unbound to localised polaronic dot states, a 
sharp polaron transition cannot occur for a dot level $\Delta \leq 0$. 
Nevertheless, we observe in figure~\ref{FigAdiab2} a transition signalled 
by the jump of $\gamma_{min}$ to $1$ with corresponding increase of $n$, 
decrease of $L$, and formation of a pronounced peak in 
$A_{dd}(\omega)$ at negative energies.
The reason is again the change of the global minimum of $E(\gamma)$, which has two local minima for larger $\varepsilon_p$. 
Since for $\Delta=0$ the interaction need not overcome a repulsive dot potential, the transition takes places at smaller $\varepsilon_p^c \approx 1.125$.
Therefore, and in contrast to the previous case, no isolated quasiparticle
peak in $A_{dd}(\omega)$ emerges at the transition, and the change of the 
spectral function is less dramatic. Above the transition the qualitative
behaviour of the imaginary part of the self-energy 
[see inset of~\ref{FigAdiab2}~(f)] is the same as for
the repulsive quantum dot  [cf. inset of~\ref{FigAdiab1}~(f)].
Since $\varepsilon_p$ is smaller now, the maxima of the 
phonon contributions to $\Sigma_{dd}$ are less separated
than in figure 4(f). 
 
In our approach the transition results from a jump in $\gamma_{min}$.
As before, this might indicate the formation of a localised 
polaronic dot state.
But we know from the various variational approaches to the polaron problem
that such jumps often arise  as artefacts of the 
variational ansatz~\cite{Feea94}. 
For the Holstein polaron with EP interaction at each lattice site, no phase transition exists~\cite{GL91}. Instead, a crossover between an almost 
free electron and a heavy polaron takes place. 
The crossover can however be very rapid for small phonon 
frequency~\cite{AFT08}.
But we also know that, in contrast to the Holstein polaron problem, 
for a single electron at a vibrating quantum dot a true phase transition, 
from $n=0$ to finite $n>0$, takes place~\cite{MNAFDCS09,AF08b,FWLB08}.
This phase transition becomes more pronounced for small $\omega_0$.
The behaviour found here therefore does not contradict the essential 
physical mechanism in our situation.
In principle, our approach mimics the sharp adiabatic polaron
transition by the change of the parameter $\gamma$ of the 
(non-adiabatic) Lang-Firsov transformation.
While the precise nature of the transition is only poorly described by this approximation, we still believe that the transition -- or rapid crossover -- 
itself is characteristic for the quantum dot at small $\omega_0$.

\subsection{Large phonon frequency}

For large phonon frequency $\omega_0=10$, in the antiadiabatic regime,
phonons adjust instantaneously to the electrons. Now our non-adiabatic 
variational Lang-Firsov approach  perfectly matches the situation.
We will see that the transitions found in the previous 
(adiabatic) cases will be replaced by smooth changes of the physical 
observables. We then note that the results obtained can be understood easily
starting from the case without EP interaction.

\subsubsection{Repulsive dot}

\begin{figure}[b]
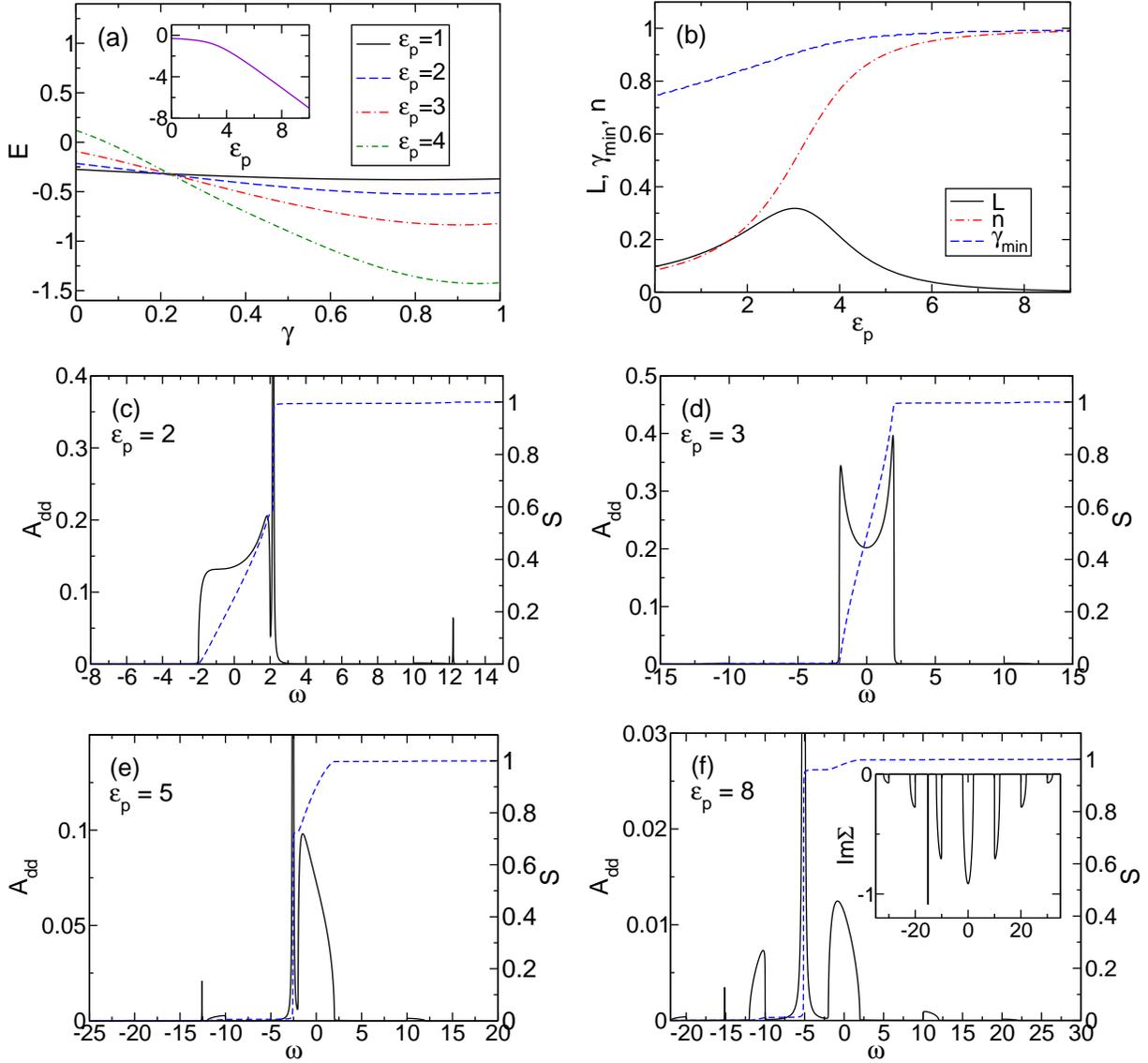

    	\includegraphics[width=0.45\textwidth]{fig5a}\hspace{1.0cm}
    	\includegraphics[width=0.44\textwidth]{fig5b}\\[0.25cm]
    	\hspace*{0.2cm}\includegraphics[width=0.49\textwidth]{fig5c}\hspace{0.3cm}
    	\includegraphics[width=0.49\textwidth]{fig5d}\\[0.25cm]
    	\includegraphics[width=0.5\textwidth]{fig5e}\hspace{0.3cm}
    	\includegraphics[width=0.5\textwidth]{fig5f}
 \caption{(Colour online) All results for $\Delta=3$, $t=t_d=1$, $\mu=0$ and $\omega_0=10$. 
Upper panels: Total energy $E$ (a) as a function of $\gamma$, for different $\varepsilon_p$ (inset: Minimum of $E$ as a function of $\varepsilon_p$) and  
optimal parameter $\gamma_{min}$, kinetic coefficient $L$ and 
particle density $n$ at the dot (b) 
as a function of $\varepsilon_p$. 
Lower panels: Dot spectral function $A_{dd}(\omega)$ and 
integrated spectral weight $S(\om)$ for different $\varepsilon_p$ (c)-(f). 
The inset in panel (f) gives the imaginary part of the dot self-energy 
$\Sigma_{dd}(\omega)$.}
 \label{FigAntiAdiab1}
\end{figure}

The fact that for large phonon frequency no transitions occur is most clearly 
seen for a repulsive barrier in figure~\ref{FigAntiAdiab1}:
All quantities depend smoothly on $\varepsilon_p$.
The total energy $E(\gamma)$ has a unique minimum for all 
$\varepsilon_p$, which is the reason why no transition occurs. 
Note that $\gamma$ grows from $\approx 0.75$ to $1$, 
as $\varepsilon_p$ is increased.
For large $\omega_0$ and $\varepsilon_p$, the Lang-Firsov transformation implements the correct physical mechanisms. Nevertheless, at weak EP coupling, the deviation $\gamma < 1$ indicates the importance of corrections to the complete Lang-Firsov transformation.

The spectral functions in figure~\ref{FigAntiAdiab1} show that, although no transition occurs, we start with a peak in $A_{dd}(\omega)$ at positive energies 
for small $\varepsilon_p$ ($<\Delta$), to end up with a polaronic 
quasiparticle signature at very strong EP coupling. 
At $\varepsilon_p=2$, the peak enters the band of lead states from above, leading 
to an asymmetric deformation of the semi-elliptic band [see panel (c)].
A second absorption feature is separated by the phonon frequency,
but carries almost no spectral weight. 
With increasing EP interaction, the effective dot level is lowered until
a ``neutral'' dot evolves at about $\tilde \Delta=\Delta-\varepsilon_p=0$,  
as can be seen from the 1D tight-binding-model-like absorption in 
panel (d) (note that we here are in the weak EP interaction 
regime since $g^2=0.3$, which is the relevant coupling 
parameter in the anti-adiabatic region, is small). At large EP coupling,
the polaronic peak appears at $\tilde \Delta<0$ and acquires a
spectral weight of nearly unity (see panel (f) for $\varepsilon_p=8$). 
Due to the large phonon frequency $\omega_0 > W$, the phononic 
sidebands do not overlap in this case (in contrast to figure~\ref{FigAdiab1}), 
and the spectral function and self-energy show the typical multi-band 
structure known from the antiadiabatic Holstein polaron. Most 
importantly, we now find intervals where 
${\rm Im} \Sigma_{dd}=0$ between the 
non-overlapping phonon (side) bands. If the polaronic peak is located   
within such an intermediate range the quasiparticle cannot decay 
by (multi-) phonon absorption or emission processes [see panel (f)].
This means the polaronic dot state acquires in principle an 
infinite lifetime (in the limit of very large couplings and 
phonon frequency). Naturally, as $\omega_0\to\infty$, we recover the 
behaviour of the impurity model, where a true bound 
state occurs~\cite{AF08b,MNAFDCS09}. 

Thus, for large phonon frequency, or whenever $\gamma$ is close to unity,  
we can understand most properties starting from the non-interacting case,
if we take the interaction into account by renormalisation 
of the appropriate physical parameters.
Inspection of equations~(\ref{LfromA}),~(\ref{tildeAfromA}) shows 
that one central effect of interaction on the kinetic coefficient
is the renormalisation of $t_d$ to an effective dot-lead hopping
$t_d e^{-\tilde{g}^2/2}$. 
The second central effect is the change in the dot density of states,
which is to a large extent caused by lowering of the effective dot energy
(below the value $\Delta$ without interaction) due to deformation
of the quantum dot in the presence of electrons.

The simple picture is valid only in the limit $\gamma=1$,
when the dot energy is effectively lowered by $-\varepsilon_p$,
such that $\tilde \Delta = \Delta-\varepsilon_p$ in equation~(\ref{tr2}),
and the dot-lead hopping is effectively reduced by $e^{-g^2/2}$,
such that $\tilde A_{dd}(0)  =  e^{- g^2} A_{dd}(0) $ in
equation~(\ref{tildeAfromA}).
The kinetic coefficient $L$ then has properties analogous to the non-interacting case, with the appropriately renormalised parameters. 
We discussed above (section~\ref{sec:NoInter}) the consequences for $L$ resulting from a change of $t_d$ or $\Delta$.
For the curve shown in figure~\ref{FigAntiAdiab1},
it turns out that it can be indeed reproduced from the expression for $L$ in the non-interacting case, evaluated with an effective dot energy 
$\Delta - \varepsilon_p$ and effective dot lead 
hopping $\tilde{t_d} = t_d e^{-g^2/2}$ replacing $\Delta$, $t_d$.
In particular, $L$ is maximal for $\varepsilon_p = \Delta$ (cf. figure~\ref{Fig2}).
Note that away from the limit of large phonon frequency, 
whenever $\gamma \ll 1$, different behaviour is found.
Also the shape of the dot spectral function, and especially the value of $A_{dd}(0)$, is modified in addition to simple renormalisation. 
Of course, and similar as for the polaron problem, the retardation of the
EP interaction manifests itself most prominently at  small to intermediate
phonon frequency.

\subsubsection{The case $\Delta =0$}

\begin{figure}[ht]
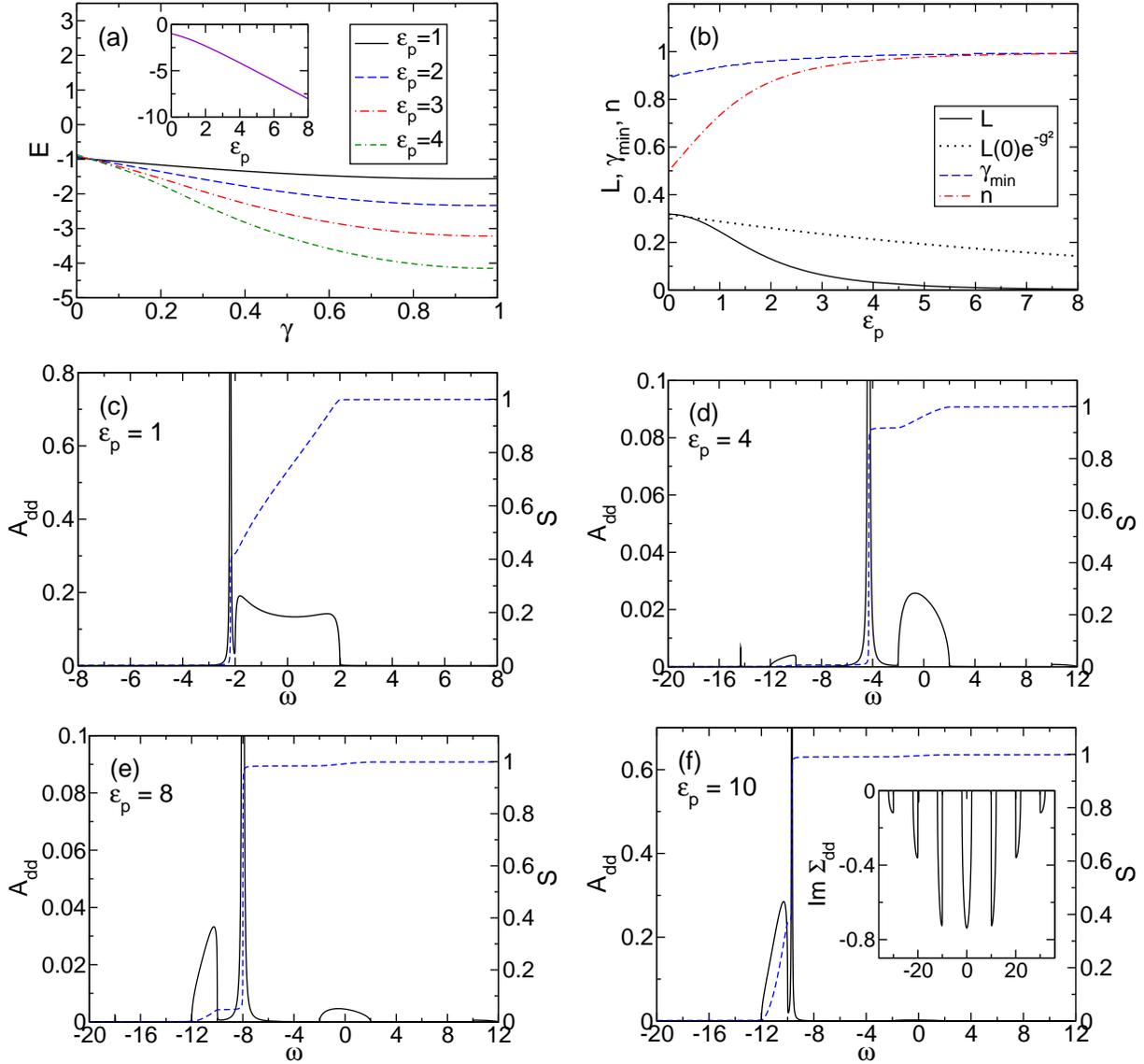

    	\hspace*{0.3cm}\includegraphics[width=0.43\textwidth]{fig6a}\hspace{1.25cm}
    	\includegraphics[width=0.44\textwidth]{fig6b}\\[0.25cm]
    	\includegraphics[width=0.5\textwidth]{fig6c}\hspace{0.3cm}
    	\includegraphics[width=0.5\textwidth]{fig6d}\\[0.25cm]
    	\includegraphics[width=0.5\textwidth]{fig6e}\hspace{0.3cm}
    	\includegraphics[width=0.5\textwidth]{fig6f}
 \caption{(Colour online) All results for $\Delta=0$, $t=t_d=1$, $\mu=0$ and $\omega_0=10$. 
Upper panels: Total energy $E$ (a) as a function of $\gamma$, for different $\varepsilon_p$ (inset: Minimum of $E$ as a function of $\varepsilon_p$) and  
optimal parameter $\gamma_{min}$, kinetic coefficient $L$ and 
particle density $n$ at the dot (b) 
as a function of $\varepsilon_p$. 
Lower panels: Dot spectral function $A_{dd}(\omega)$ and 
integrated spectral weight $S(\om)$ for different $\varepsilon_p$ (c)-(f). 
The inset in panel (f) gives the imaginary part of the dot self-energy 
$\Sigma_{dd}(\omega)$.}
 \label{FigAntiAdiab2}
\end{figure}

The behaviour for the attractive dot is similar to the previous 
case (see figure~\ref{FigAntiAdiab2}).
Here, of course, a pronounced peak in $A_{dd}(\omega)$ occurs 
at negative energies for all $\varepsilon_p$. 
Once again, all features can be understood starting from 
the non-interacting case with appropriate renormalisation, as explained above.
Since $\tilde \Delta \leq 0$ for all $\varepsilon_p \geq 0$, the kinetic 
coefficient has no maximum as a function of $\varepsilon_p$.

The simple picture given above takes into account only 
the renormalisation of $t_d$ and $\Delta$. 
It is important to keep in mind that both effects 
lead to a reduction of $L$.
As a consequence, the change of the kinetic coefficient is not simply 
given by an exponential behaviour $\propto e^{-g^2}$ (compare $L$ to 
the dashed curve in figure~\ref{FigAntiAdiab2}).
In the present case, the coupling strength is small in terms of the average number of phonons $g^2 = \varepsilon_p / \omega_0$, for which $g^2 < 1$,
but large in terms of the shift of the dot energy $\varepsilon_p$, which is of the order of the bandwidth $W$.
Here, the reduction of $L$ is mainly caused by this large shift.

The opposite situation can occur for small phonon frequency, when
$g^2$ is large already for small $\varepsilon_p$.
Then, however, the renormalisation of $t_d$ is not adequately described by an exponential factor $e^{-g^2/2}$.
In the limit $\omega_0 \to 0$ of small phonon frequency, 
$g^2 = \varepsilon_p / \omega_0 \to \infty$ for any $\varepsilon_p > 0$.
If the exponential dependence $\propto e^{-g^2}$ persisted,
that would imply zero current even for tiny $\varepsilon_p$, which is unphysical.
A calculation with fixed $\gamma=1$ therefore overestimates the reduction of $L$ for intermediate-to-small phonon frequencies.
We discussed in the previous subsection how, in our treatment, variation of the parameter $\gamma$ accounts partially for this deviation,
leading to $\gamma  \ll 1$ away from the antiadiabatic strong-coupling limit.

\subsection{Intermediate phonon frequency}

\begin{figure}[t]
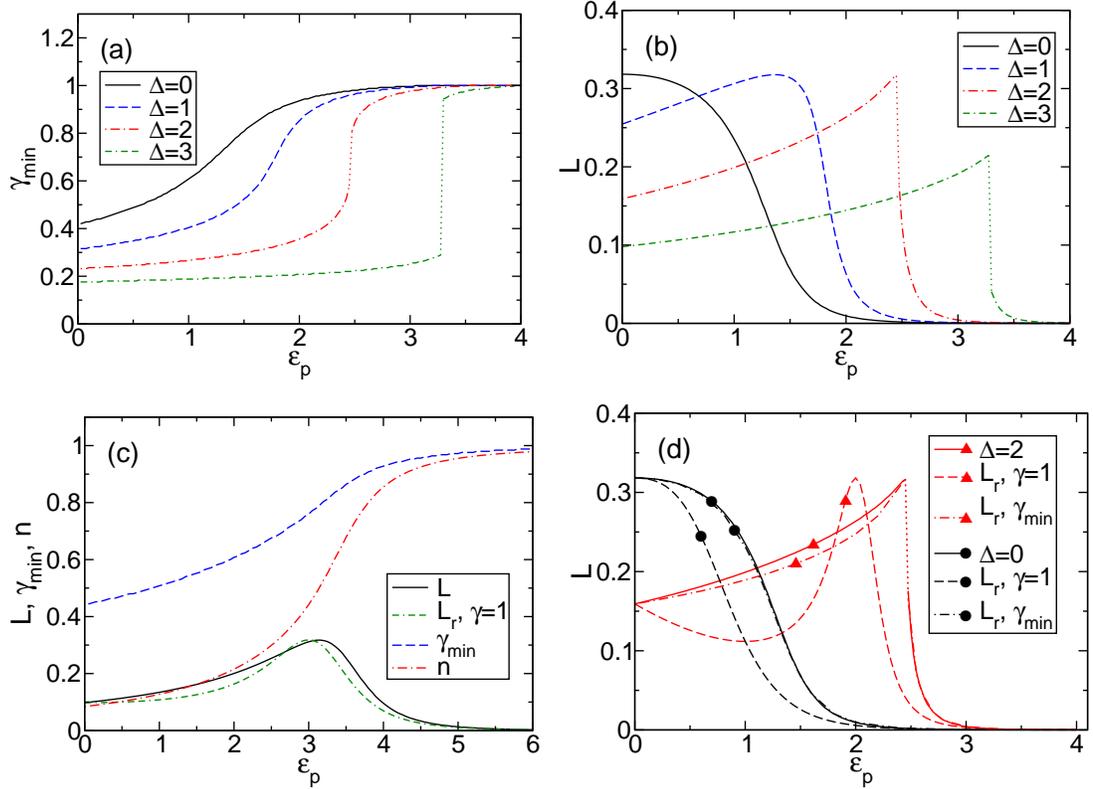

\hspace*{2cm}\includegraphics[width=0.44\textwidth]{fig8a}\hspace{0.3cm}
\includegraphics[width=0.44\textwidth]{fig8b}\\[0.25cm]
\hspace*{2cm}\includegraphics[width=0.45\textwidth]{fig8c}\hspace{0.3cm}
\includegraphics[width=0.44\textwidth]{fig8d}\\[0.25cm]
\caption{(Colour online) All results for $t=t_d=1$, $\mu=0$.
Upper row: For $\omega_0=1$, optimal $\gamma_{min}$ (a) and 
kinetic coefficient $L$ (b) for various $\Delta$ as indicated.
Lower row, panel (c): For $\Delta=3$ and $\omega_0=3$, optimal parameter 
$\gamma_{min}$, kinetic coefficient $L$, and
particle density $n$ at the dot, as a function of $\varepsilon_p$. 
The dashed green curve shows $L$ calculated for the non-interacting case with renormalised parameters (see text).
Lower row, panel (d): For $\omega_0=1$ and two different $\Delta$, kinetic coefficient $L$ as a function of $\varepsilon_p$.
The dashed curves show $L$ calculated for the non-interacting case with renormalised parameters, but fixed $\gamma=1$. 
The dot-dashed curves have been obtained taking the parameter $\gamma$ from the upper left panel (see text).

 }
\label{FigInter}
\end{figure}

For intermediate phonon frequencies the qualitative behaviour depends crucially on the value of $\gamma$, even if no transition occurs.
From our previous discussion we know that both a positive $\Delta$ or a small $\omega_0$ favour a rapid, or even discontinuous, transition.
For $\omega_0=1$, we show in figure~\ref{FigInter} (upper row) how a smooth crossover evolves into a sudden transition with increasing $\Delta$.
In contrast to the case of small phonon frequency $\omega_0=0.1$, the kinetic coefficient $L$ is a smooth function of $\varepsilon_p$ for $\Delta=0$.
A transition in $L$ occurs only for larger $\Delta$.
Increasing the phonon frequency to $\omega_0=3$ [lower row, panel (c)] 
then leads again to a smooth crossover even at $\Delta=3$.

Changing the phonon frequency, 
we should ask to which extent the renormalisation scenario given for the antiadiabatic case remains applicable.
For $\omega_0=3$ [panel (c)] we observe that $L$ differs from the value obtained, as in the previous subsection, from the non-interacting case for renormalised $\Delta$, $t_d$ (in particular the maximum of $L$ occurs for $\varepsilon_p > \Delta$),
but although $\Delta=3$, the two curves match rather well.
The situation changes for $\omega_0=1$ [panel (d)],
where strong deviations occur already for $\Delta=0$
(note that the dashed curve for $\Delta=2$ even misses the 
increase of $L$ at smaller $\varepsilon_p$).
Evidently, the simple renormalisation scenario fails, as we expected.
We can achieve much better agreement if we perform the same 
calculation but incorporate the parameter $\gamma$ taken from 
the upper left panel in figure~\ref{FigInter} 
(the dashed curves would correspond to fixed $\gamma=1$).
Small deviations remain for $\Delta=2$, since the full calculation 
includes damping of states, indicated by a finite imaginary part of 
the self-energy, which is not captured by the change of $\gamma$.

It is now evident that the essential feature of our calculation is the 
self-consistent determination of the parameter $\gamma$.
Once we know its value, we may get a good approximation already with a 
modified renormalisation argument which was originally constructed for 
the antiadiabatic limit.
If, in contrast, we fix $\gamma=1$ we will miss the physics away from the 
limit of large phonon frequencies.
The restricted use of the Lang-Firsov transformation for 
intermediate-to-small phonon frequencies is well known in 
the Holstein polaron literature.
It is important to realize that this restriction applies 
also to the situation of a vibrating quantum dot.

\subsection{Variation of the chemical potential}
So far all results were given for chemical potential $\mu=0$.
A change of the chemical potential affects the kinetic coefficient 
in two ways. First, since in equation~(\ref{LfromA}) the lead 
density of states $\varrho(\xi)$ and the dot spectral function 
$A_{dd}(\omega)$ are evaluated at the chemical potential,
a change of $\mu$ results in a change of $L$. 
%
Second, phonon emission/absorption is possible only if 
free states are accessible after an electron changed its 
energy by $\pm s\omega_0$. Otherwise, EP interaction is suppressed 
by Pauli blocking. Therefore, the shape of $A_{dd}(\omega)$ itself 
does depend on $\mu$ in a true many-particle calculation as performed here.
Significant changes occur whenever $\pm \omega_0$ crosses the 
band edges (at about $\pm W-\mu $ at weak coupling).
\begin{figure}[t]
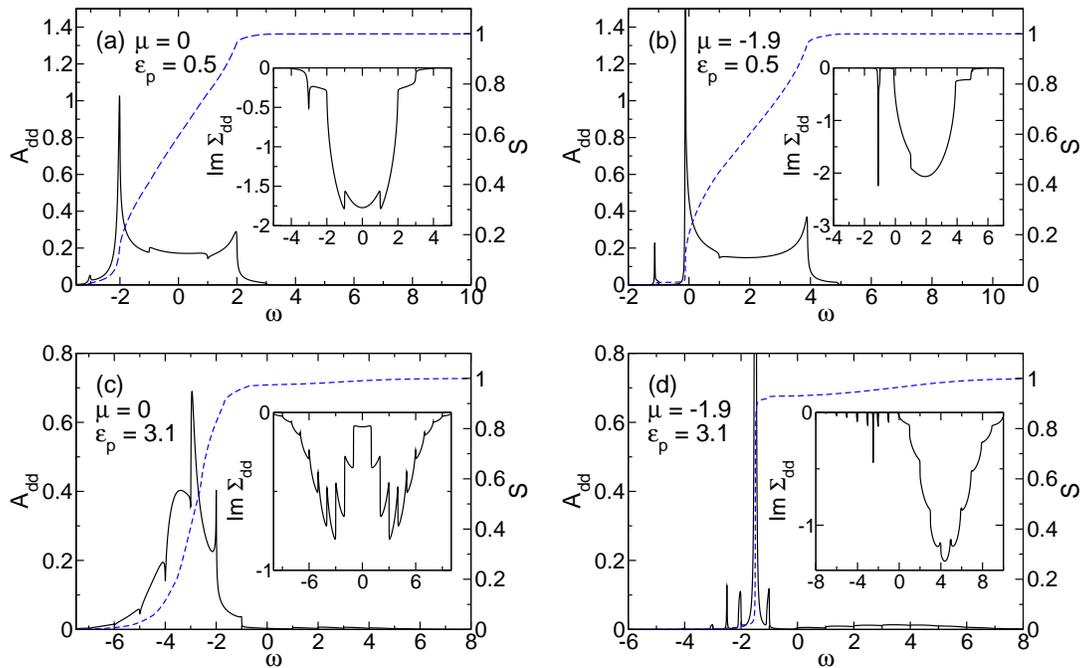

 \hspace*{1cm} \includegraphics[width=0.44\textwidth]{fig9a}\hspace{0.3cm}
  \includegraphics[width=0.44\textwidth]{fig9b}\\[0.25cm]
 \hspace*{1cm} \includegraphics[width=0.44\textwidth]{fig9c}\hspace{0.3cm}
  \includegraphics[width=0.44\textwidth]{fig9d}
  \caption{(Colour online) All results for $\Delta=0$, $t=t_d=1$ and 
$\omega_0=1$. 
Comparison of $A_{dd}(\omega)$, $S(\om)$, and 
the self-energy $\Sigma_{dd}(\omega)$ (insets) 
for $\mu=0$ [left column, panels (a) and (c)] 
and $\mu=-1.9$ [right column, panels (b) and (d)],
for weak coupling $\varepsilon_p=0.5$ [upper row, panels (a) and (b)] 
and strong coupling $\varepsilon_p=3.1$ [lower row, panels (c) and (d)].
}
\label{FigChem1}
\end{figure}

\begin{figure}[ht]
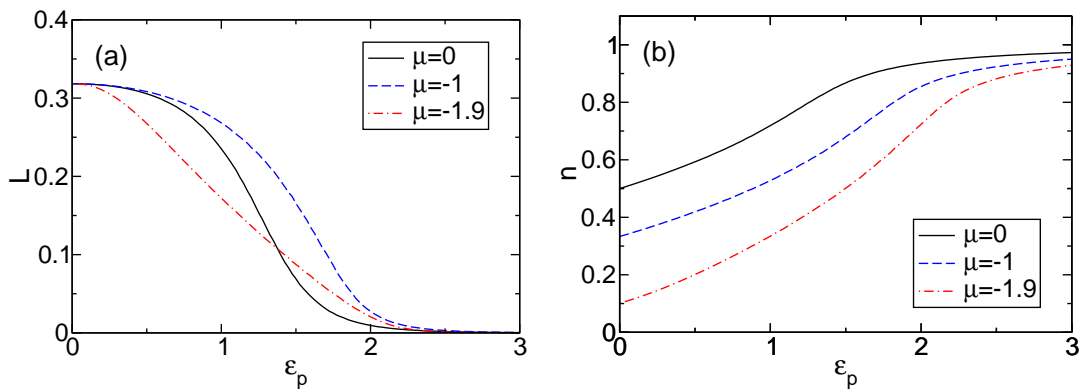

 \hspace*{1cm}\includegraphics[width=0.44\textwidth]{fig10a}\hspace{0.3cm}
 \includegraphics[width=0.44\textwidth]{fig10b}\hspace{0.3cm}
\caption{(Colour online) All results for $\Delta=0$, $t=t_d=1$ 
and $\omega_0=1$. Kinetic coefficient $L$ [panel (a)] and dot density of 
states $n$ [panel (b)] as a function of $\varepsilon_p$, for varying 
chemical potential $\mu$.}
\label{FigChem2}
\end{figure}

This effect is evident in the spectral function $A_{dd}(\omega)$ in figure~\ref{FigChem1}.
At weak coupling ($\varepsilon_p=0.5$) the shape of $A_{dd}(\omega)$ is similar for half-filling ($\mu=0$) and small particle density ($\mu=-1.9$), 
but small differences at the lower band edge are a first indication of the different behaviour at stronger coupling.
There, for $\varepsilon_p=3.1$, the spectrum for $\mu=0$ is completely incoherent,
with finite $\Im \Sigma_{dd}(\omega)$. 
Around $\omega=0$ we observe a valley in $\Im \Sigma_{dd}(\omega)$ of width $2\omega_0$, which results from Pauli blocking of states in the vicinity of the Fermi energy (cf. the discussion in Ref.~\cite{LHF06}).
Note that $\Im \Sigma_{dd}(\omega) \ne 0$ here even at the Fermi energy, 
since the self-energy contains the contribution from dot-lead transfer.
For $\mu=-1.9$, states below the phonon emission threshold, located $\omega_0$ above the lower band edge, cannot emit a phonon (phonon absorption is suppressed at zero temperature). Electrons in these states are undamped, with infinite lifetime corresponding to $\Im \Sigma_{dd}(\omega)=0$.

The interpretation of the behaviour of the kinetic coefficient $L$ (see figure~\ref{FigChem2}) relies on these two mechanisms.
First, if $\mu$ decreases, the change in the density of states should reduce the value of $L$ [compare the curves for $\mu=0$ (solid line) 
and $\mu=-1.9$ (dot-dashed line)].
Also, the dot density of states $n$ decreases.
We note that for the non-interacting $(\varepsilon_p=0)$ 1D case the changes in $\varrho(\xi)$ and $A_{dd}(\omega)$ cancel by chance, and $L$ is independent of $\mu$.
However, at stronger coupling, the different influence of Pauli blocking reverses this behaviour, and $L$ is larger for smaller $\mu$.
This explains why the curve for $\mu=-1.9$ crosses the curve for $\mu=0$ in figure~\ref{FigChem2}.

\subsection{Small dot-lead hopping (tunnel contacts)}

We have so far discussed the importance of the phonon frequency 
only in the situation $t=t_d$.
On physical grounds it is the ratio $\omega_0 / t_d$, instead 
of $\omega_0 / t$, which should distinguish the adiabatic from 
the antiadiabatic regime.

\begin{figure}[ht]
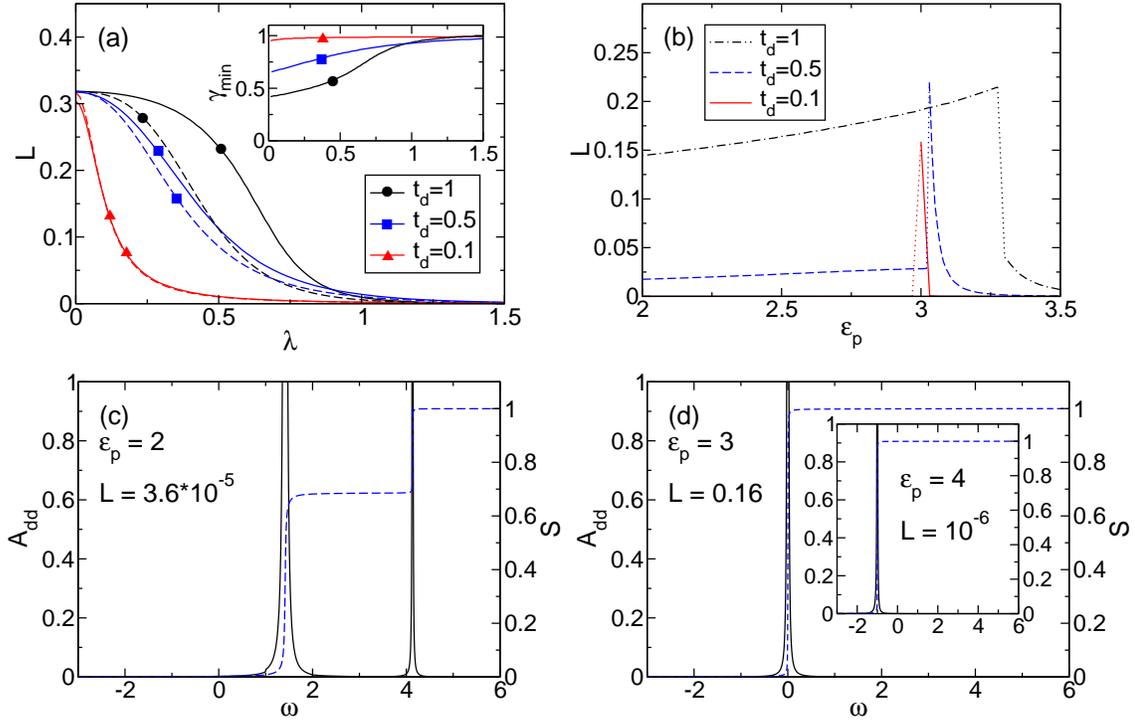

 \hspace*{1.05cm} \includegraphics[width=0.43\textwidth]{fig7a}\hspace{0.5cm}
  \includegraphics[width=0.43\textwidth]{fig7b}\\[0.25cm]
 \hspace*{1cm} \includegraphics[width=0.47\textwidth]{fig7c}\hspace{0.1cm}
  \includegraphics[width=0.47\textwidth]{fig7d}
\caption{(Colour online) All results for $t=1$, $\mu=0$ 
and $\omega_0=1$.
Upper row: Kinetic coefficient $L$ as a function of 
$\lambda=\varepsilon_p/2t_d$ at $\Delta=0$ [panel (a)] and 
as a function of $\varepsilon_p$ at $\Delta=3$
[panel (b)] for different $t_d$. 
In panel (a) $L$ is compared to the  renormalised non-interacting
case (dashed lines, see text). 
The inset in panel (a) gives the corresponding 
optimal parameter $\gamma_{min}$. 
Lower row: Dot spectral function $A_{dd}(\om)$ and integrated 
spectral weight $S(\om)$ at $\Delta=3$, $t_d=0.1$ for 
$\varepsilon_p=2$ (c), $\varepsilon_p=3$ (d), and $\varepsilon_p=4$
[inset, panel (d)].}
\label{Figtd}
\end{figure}

In figure~\ref{Figtd} (a) we show, for intermediate phonon frequency 
$\omega_0=1$ and $\Delta=0$, the change of behaviour as $t_d$ is 
reduced by one order of magnitude. The kinetic coefficient $L$ decreases with $t_d$ (see upper panels). Note that if  we reduce $t_d$
at fixed $\lambda=\varepsilon_p / (2t_d)$ in  figure~\ref{Figtd}~(a), 
we de facto reduce the coupling $\varepsilon_p$, but nevertheless 
$L$ decreases. 
This implies that the effect of smaller $t_d$ on $L$ dominates over 
the possible increase of $L$ for smaller coupling.
We compare $L$ again with the value obtained in the renormalisation 
scenario, with a calculation as in the non-interacting case but with 
renormalised parameters $\Delta$, $t_d$ (see dashed lines). 
For $t_d=t=1$, and $\omega_0/t_d=1$, both curves disagree.
We already discussed above that this is a consequence of 
adiabatic corrections, which become important at 
intermediate-to-small phonon frequency. However, if we reduce 
$t_d$ to $0.1$ while keeping $\omega_0=1$ fixed, and thereby 
increase $\omega_0/t_d$ to $10$, both curves match.
Apparently, we enter the antiadiabatic regime by sufficient 
reduction of $t_d$. This indicates that indeed $\omega_0/t_d$ 
is the relevant ratio to distinguish the adiabatic from the 
antiadiabatic regime. For $\omega_0 / t_d \gg 1$, and independent 
of $\omega_0/t$, the system has physical properties that can be 
described within the simple renormalisation scenario associated 
with the complete Lang-Firsov transformation. 

In  figure~\ref{Figtd}~(b) we make the same observation for 
$\Delta=3$, still with $\omega_0=1$.
For $t_d \gtrsim 0.5$ the kinetic coefficient $L$ shows the 
transition familiar to us from the previous discussions of 
intermediate or small phonon frequencies, which is in contrast 
to the physics in the antiadiabatic regime.
For $t_d \ll 1$ a sharp peak  occurs in $L$ for $\varepsilon_p = \Delta$.
This is of course the behaviour expected for the antiadiabatic regime, 
which is reminiscent of the non-interacting case for small $t_d$ 
with a peak of $L$ at $\Delta=0$ (cf. figure~\ref{Fig2}).

The lower panels of figure~\ref{Figtd} show the spectral function of  
the repulsive quantum dot at small dot-lead hopping $t_d=0.1$  for
$\tilde\Delta>0$~[panel (c)], $\tilde\Delta\simeq 0$ ~[panel (d)], 
and  $\tilde\Delta<0$~[inset panel (d)]. Below the ``critical'' EP coupling
we have $\gamma_{min}\simeq 0.2$ and obtain  a double-peak structure 
of $A_{dd}$ because both the first and the third term in 
equation~(\ref{sigmatwocomplex}) give significant contributions. 
At $\varepsilon_p=3$, the prefactor of the third term vanishes
($\gamma_{min}=1$), and a  single-peak structure develops. 
This polaronic peak is located at the Fermi energy and contains all 
the spectral weight. Therefore $L$ is enlarged more than three orders of
magnitude. Increasing $\varepsilon_p$ further the polaronic signal is 
narrowed and shifted away from the Fermi level. As a result
$L$ decreases off by five orders of magnitude.

\begin{figure}
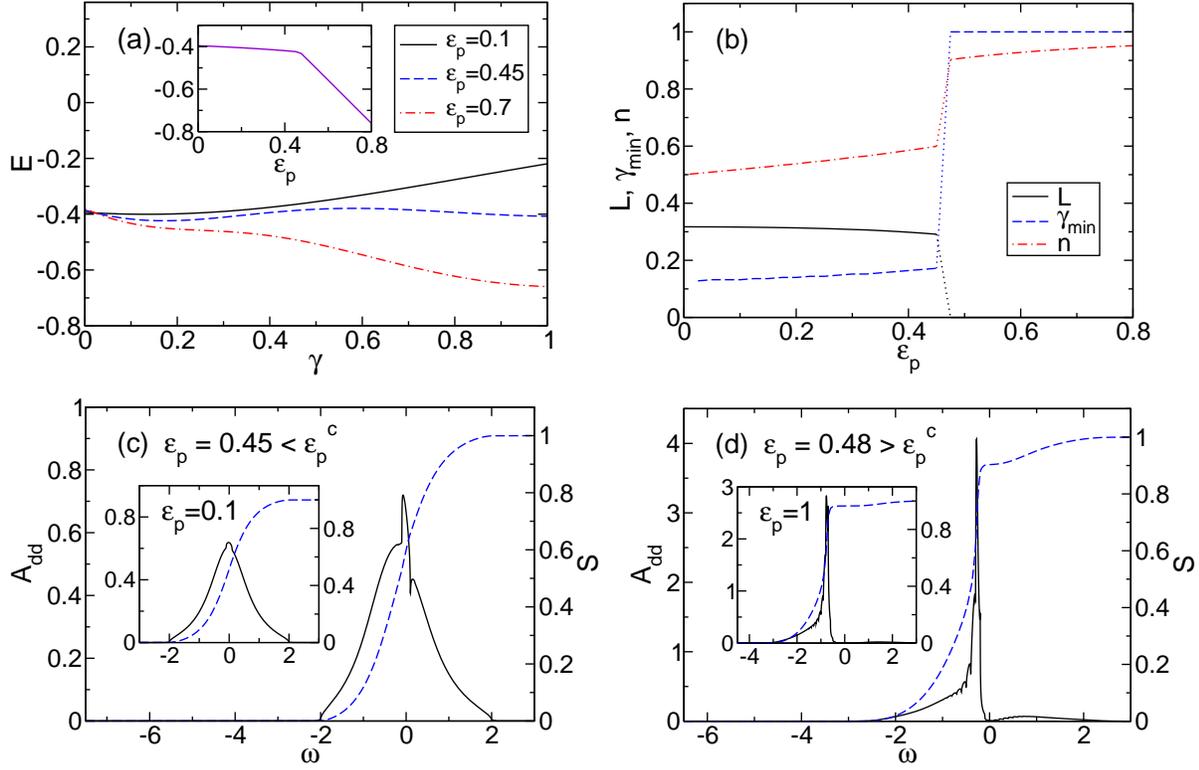

    	\includegraphics[width=0.46\textwidth]{fig11a}\hspace{0.6cm}
    	\includegraphics[width=0.46\textwidth]{fig11b}\\[0.25cm]
    	\includegraphics[width=0.5\textwidth]{fig11c}\hspace{0.3cm}
    	\includegraphics[width=0.48\textwidth]{fig11d}
 \caption{(Colour online) All results for $t=1$, $\Delta=0$, $t_d=0.5$, $\mu=0$, 
and $\omega_0=0.1$. Panel (a): $E$ as a function of $\gamma$ for different 
$\varepsilon_p$ (inset: minimum of $E$ as a function of $\varepsilon_p$). 
Panel (b): $L$, $\gamma_{min}$, and $n$ as functions 
of $\varepsilon_p$. For $\varepsilon_p^c\approx0.46$ a crossover 
takes place. Panels (c) and (d) Spectral functions $A_{dd}(\om)$ and integrated 
spectral weight $S(\om)$ for coupling strengths below and above 
$\varepsilon_p^c$.}
 \label{fig_adiab_smalltd}
\end{figure}
%

\begin{figure}[t]
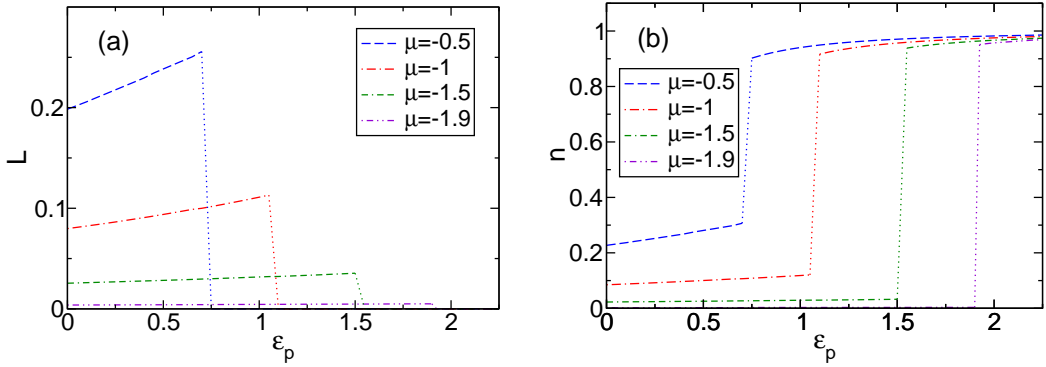

 \hspace*{1cm} \includegraphics[width=0.42\textwidth]{fig13a}\hspace{0.5cm}
    	\includegraphics[width=0.42\textwidth]{fig13b}
 \caption{
All results for $\Delta=0$, $t=1$, $t_d=0.5$ and $\omega_0=0.1$. 
Kinetic coefficient $L$ (a) and particle density on the dot $n$ (b) as functions 
of $\varepsilon_p$ for varying chemical potential $\mu$.}
 \label{fig_inter_mu1}
\end{figure}
Figure~\ref{fig_adiab_smalltd} gives more results for the 
experimentally relevant wide-band case, $t>t_d$, now
in the adiabatic regime. Owing to the values of the
parameters $\mu$ and $\om_0$, we have a situation 
where $\Im \Sigma_{dd}(\om)\neq 0$ in the whole relevant $\om$ region 
and the spectral function is given by equation~(\ref{inc_spectrum}),
According to the formula for $\Im \Sigma_{dd}(\om)$, 
equation~(\ref{im_sigma}), the cases $\varepsilon_p=0.1$ 
and $\varepsilon_p\lesssim \varepsilon_p^c$
 [panel~(c)] show the predominance of the first term. The shift of spectral weight to negative $\om$ becomes apparent 
for $\varepsilon_p=0.45$, indicating the influence of the EP
interaction. The spectral functions for  $\varepsilon_p\gtrsim 
\varepsilon_p^c$  and $\varepsilon_p=1$ [panel~(d)] make 
evident the suppression of the first-term contribution and the 
multi-phonon structure according to the second term in 
equation~(\ref{im_sigma}). The maxima of the spectral functions are situated near $\om =\tilde\Delta$.
Then again the sudden decrease of $L$ at $\varepsilon_p^c$
may be understood from equation~(\ref{tildeAfromA}) 
by the sudden change of $A_{dd}(0)$ at $\varepsilon_p^c$.

\begin{figure}[t]
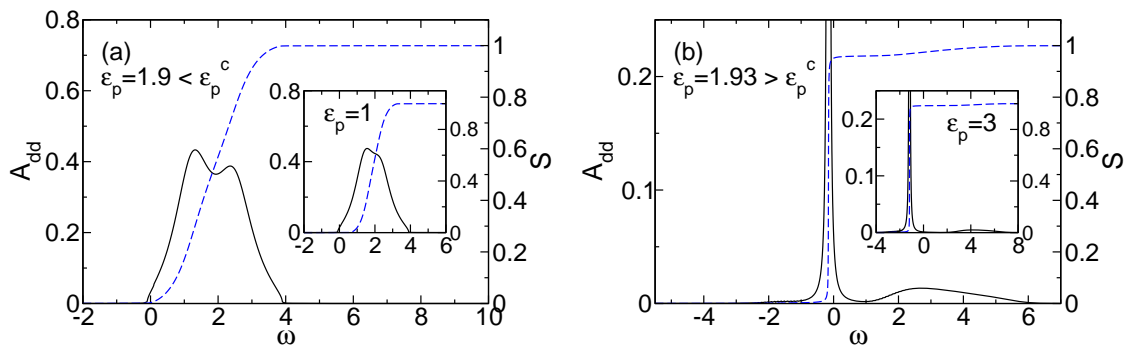

\hspace*{1cm} \includegraphics[width=0.46\textwidth]{fig14a}\hspace{0.3cm}
    	\includegraphics[width=0.46\textwidth]{fig14b}
 \caption{(Colour online) Dot spectral function for $\Delta=0$, $t=1$, $t_d=0.5$, $\omega_0=0.1$, $\mu=-1.9$ and varying $\varepsilon_p$.}
 \label{Fig_inter_mu2}
\end{figure}
Figure~\ref{fig_inter_mu1}, for $\omega_0=0.1$, shows the kinetic coefficient $L$ 
and the particle density on the dot, $n$, for $\mu < 0$, 
whereas the dot spectral function is given in 
figure~\ref{Fig_inter_mu2} for $\mu=-1.9$ only. 
Again, we observe an adiabatic transition in $L$ and $n$.
As $\mu$ decreases, the critical EP-coupling strength moves 
to larger values, simply because
the effective dot level has to be lowered by a larger
$\varepsilon_p$ to roughly match the Fermi level. 
%
%
\begin{figure}[t]
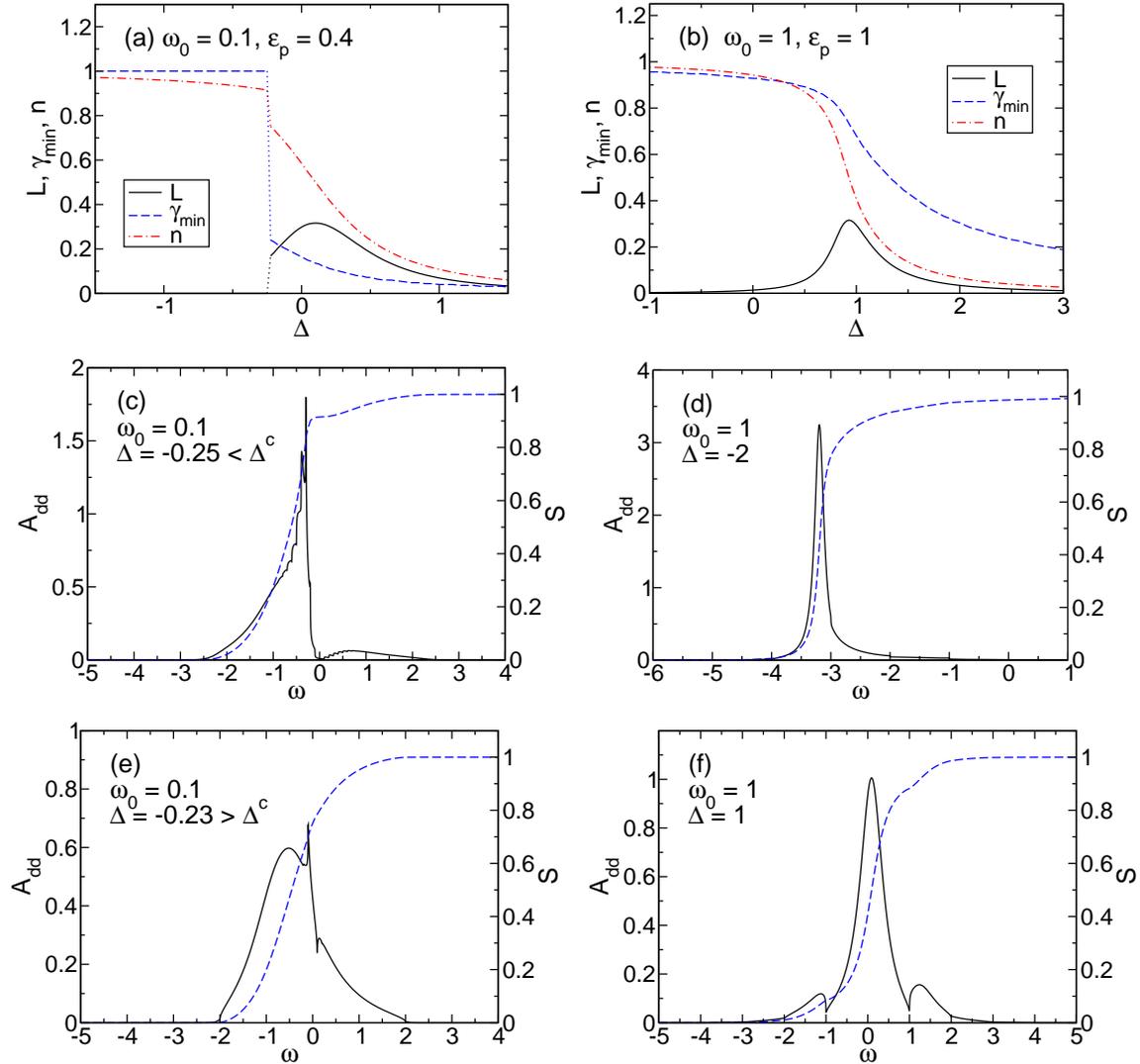

\hspace*{.9cm}\includegraphics[width=0.415\textwidth]{fig12a}\hspace{.8cm}
      	\includegraphics[width=0.42\textwidth]{fig12b}\\[0.25cm]
   \hspace*{.7cm} \includegraphics[width=0.47\textwidth]{fig12c}\hspace{0.4cm}
    	\includegraphics[width=0.45\textwidth]{fig12d}\\[0.25cm]
    \hspace*{.7cm}\includegraphics[width=0.47\textwidth]{fig12e}\hspace{0.3cm}
    	\includegraphics[width=0.47\textwidth]{fig12f}
 \caption{(Colour online) Left column: $L$, $\gamma_{min}$, and $n$ 
 as functions 
of $\Delta$ for $t=1$, $t_d=0.5$, $\omega_0=0.1$, $\varepsilon_p=0.4$ 
and $\mu=0$ [(panel (a)]. At $\Delta^c\approx -0.24$ a crossover 
takes place. Spectral functions for $\Delta$ below [panel (c)] and above 
$\Delta^c$ [panel (e)]. Right column: 
$L$, $\gamma_{min}$ and $n$ as functions of $\Delta$ 
for $t_d=0.5$, $\omega_0=1$, $\varepsilon_p=1$ and $\mu=0$
[panel (b)]. Panels (d) and (f) give the spectral functions $A_{dd}(\om)$ 
and integrated spectral weight $S(\om)$ for various $\Delta$.}
 \label{Fig_delta}
\end{figure}
For $\varepsilon_p<\varepsilon_p^c$  
[panel~(a)], the spectrum lies in the interval 
$[-W-\mu,W-\mu]\simeq[-0.1,3.9]$,
where the first term of equation~(\ref{im_sigma}) contributes,
with the apparent influence of EP coupling. For
 $\varepsilon_p>\varepsilon_p^c$ [panel~(b)], the spectral weight is shifted to a pronounced peak below the Fermi level. Because $\om_0 \lesssim  W-|\mu|$, we find no intervals where $\Im\Sigma_{dd}(\om)=0$.
%
%
%

Finally, we monitor for the wide-band case the transition 
induced by an increasing dot level $\Delta$ (see figure~\ref{Fig_delta}
at small ($\om_0=0.1$, left-hand column) and intermediate--to-large 
($\om_0=1$, right-hand column) phonon frequencies). In both cases
$\Im \Sigma_{dd}(\om)\neq 0$ for all $\om$. As discussed above, for 
$\Delta \neq 0$, the maximum in $A_{dd}$ occurs near 
$\om=\tilde{\Delta}=\Delta-\varepsilon_p\gamma(2-\gamma)$. 
In particular, for $\Delta=-2$ (panel~(d))
the spectrum consists practically only of one peak 
at about $\Delta-\varepsilon_p$ with relatively small
linewidth. Hence  only a weakly damped localised state
of the current carrier on the dot exists, having an
energy lowering equal to $\varepsilon_p$. The transition from
a localised to a delocalised carrier is accompanied by the shift 
of spectral weight to larger frequencies and the influence
of the first term in equation~(\ref{im_sigma}) is recovered.
The change of $A_{dd}(0)$ with $\Delta$ leads to the maximum
observed for $L$ in panel~(b).
Because the optimal variational parameter $\gamma_{min}$ is a continuous function of $\Delta$ with a wide range of values (panel~(b)), the effective renormalisation of $\tilde{t}_d$ and $\tilde \Delta$ depends on the dot level $\Delta$ itself. In contrast to the result for a complete Lang-Firsov transformation with fixed $\gamma=1$ 
(cf. figure 5 of~\cite{GNR06}), we therefore find a shift of the maximum of $L(\Delta)$ by less than $\varepsilon_p$ and $L$ decreases asymmetrically away from this point. However, in accordance with~\cite{GNR06}, we find no phonon side band in $L(\Delta)$.

\section{Summary}\label{sec:conclusion}

In this work, we have presented an approach to transport through a vibrating molecular quantum dot, which extends a previously developed description for the many-polaron problem.
The virtue of this approach lies in an incomplete variational Lang-Firsov transformation in which the degree of the transformation is determined self-consistently.
In this way, our approach can describe polaronic effects on transport away 
from the strong-coupling anti-adiabatic regime.
Descriptions based on a full Lang-Firsov transformed Hamiltonian are, 
in contrast, restricted to this limit.

With our approach we studied the molecular quantum dot in different regimes, 
from weak to strong coupling and small to large phonon frequency.
The dot spectral-functions, calculated within a second-order 
equation of motion approach, allow for a detailed analysis of the 
dynamical properties of the quantum dot in dependence of the model parameters. 
 Our results show that the use of an incomplete Lang-Firsov transformation
is essential to capture the physics for all but very large phonon frequencies:
in many cases the optimal parameter $\gamma$ differs significantly from unity. 

The present study is open for extension in several important directions.
On the one hand, extension to finite voltage bias is necessary. Since our approach is developed in the Green function formalism, this extension, e.g. using Keldysh techniques, is possible and will be addressed next.
On the other hand, our approach correctly captures the physics for a large range of possible parameters, but even with an incomplete Lang-Firsov transformation one encounters problems at very small phonon frequency. 

In conclusion, the presented work carries over important concepts and ideas well known from polaron physics, especially the crucial modification of the Lang-Firsov transformation, to the study of vibrating molecular quantum dots.

\section*{Acknowledgements}
This work was supported by Academy of Sciences Czech Republic 
(J.L.), Deutsche Forschungsgemeinschaft through SFB 652 (A.A.), 
and U.S. Department of Energy (A.R.B.). H.F. acknowledges the
hospitality at the Institute of Physics ASCR and Los Alamos 
National Laboratory. The authors would like to thank M. Hohenadler
and G. Wellein for valuable discussions.

\section*{References}
\bibliographystyle{jphysicsB}

 \end{document}